\documentstyle[nato,epsfig,rotate,axodraw,graphicx]{crckapb}

\begin{opening}

\title{B PHYSICS AND CP VIOLATION}

\author{R. FLEISCHER}

\institute{Deutsches Elektronen-Synchrotron DESY\\
Notkestr.\ 85, D--22607 Hamburg, Germany}

\end{opening}

\runningtitle{B PHYSICS AND CP VIOLATION}

\begin{document}

% The \begin{document} command comes after the \end{opening}
% command.

\begin{abstract}
After an introduction to the Standard-Model description of $CP$ violation
and a brief look at the present status of this phenomenon in the kaon 
system, a classification of non-leptonic $B$-decays is given and the 
formalism of $B^0_{d,s}$--$\overline{B^0_{d,s}}$ mixing is discussed. 
We then turn to the $B$-factory benchmark modes, $CP$ violation in charged 
$B$ decays, and the $B_s$-meson system. Finally, we focus both on 
$B\to\pi K$ decays, which play an important role to probe the CKM angle
$\gamma$, and on the $B_d\to\pi^+\pi^-$, $B_s\to K^+K^-$ system, which 
allows an interesting determination of $\beta$ and $\gamma$. 
\end{abstract}

\vspace*{-11.5truecm}
\begin{flushright}
\begin{tabular}{l}
DESY 00--170\\
hep--ph/0011323\\
November 2000
\end{tabular}
\end{flushright}
\vspace*{8.9truecm}
\begin{center}
{\small{\it (Invited lecture at NATO ASI 2000, Cascais, Portugal, 
26 June -- 7 July, 2000)}}
\end{center}
\vspace*{-0.2truecm}

\section{Introduction}
The violation of the $CP$ symmetry, where $C$ and $P$ denote the 
charge-conjugation and parity-transformation operators, respectively,
is one of the fundamental phenomena in particle physics. 
Although weak interactions are neither invariant under $P$, nor invariant 
under $C$, it was originally believed that the product $CP$ was preserved.
Consider, for instance, the process
\begin{equation}
\pi^+\to e^+\nu_e~\stackrel{C}{\longrightarrow}~\pi^-\to 
e^-\nu_e^C~\stackrel{P}{\longrightarrow}~\pi^-\to e^-\overline{\nu}_e.
\end{equation}
Here the left-handed $\nu_e^C$ state is not observed in nature; only after
performing an additional parity transformation we obtain the right-handed
electron antineutrino. In 1964, it was then found experimentally through
the observation of $K_{\rm L}\to \pi^+\pi^-$ decays that weak interactions 
are {\it not} invariant under $CP$ transformations~\cite{CP-discovery}. 
So far, $CP$ violation has only been observed in the kaon system,
and we still have few experimental insights into this phenomenon. However, 
the measurement of $CP$ asymmetries should also be ``around the corner'' 
in $B$ decays \cite{B-revs}, which are currently explored in great detail
at the $B$-factories. For a collection of basic references on $CP$ violation
and $B$ physics, the reader is referred to Refs.~\cite{basics,BF-rev}.

Studies of $CP$-violating effects are very exciting, since physics beyond 
the Standard Model is usually associated with new sources for $CP$ 
violation. Important examples are non-minimal supersymmetry, 
left-right-symmetric models, models with extended Higgs sectors, and
many other scenarios for ``new'' physics \cite{new-phys}. In this context, 
it is also interesting to note that the evidence for neutrino masses we 
got during the recent years points towards physics beyond the 
Standard Model \cite{akhmedov}, raising the question of $CP$ violation in the
neutrino sector \cite{lindner}, which may be studied in the more distant 
future at $\nu$-factories. In cosmology, $CP$ violation plays also a
crucial role: one of the necessary conditions to generate the 
matter--antimatter asymmetry of our Universe is -- in addition to baryon 
number violation and deviations from thermal equilibrium -- that the 
elementary interactions have to violate $CP$ (and $C$) 
\cite{sakharov,buchmueller}. Recent model calculations indicate, however, 
that the $CP$ violation present in the Standard Model is too small to 
generate the observed matter--antimatter asymmetry of ${\cal O}(10^{-10})$ 
\cite{shapos}. 

Concerning quantitative tests of the Standard-Model description of $CP$ 
violation, the $B$-meson system is particularly promising. In the 
search for new physics, it is crucial to have $CP$-violating $B$-decay
processes available that can be analysed reliably within the framework
of the Standard Model, which will be the major topic of these lectures. 
Presently, we are at the beginning of the $B$-factory 
era in particle physics, and in the summer of 2000, the BaBar (SLAC) and 
Belle (KEK) collaborations have already reported their first results. 
Moreover, HERA-B (DESY) has seen first events, CLEO-III (Cornell) has 
started taking data, and run II of the Tevatron (Fermilab) will follow next 
spring. A lot of interesting physics will also be left for 
``second-generation'' $B$-experiments at hadron machines, LHCb (CERN) 
and BTeV (Fermilab). Detailed studies of the $B$-physics potentials of 
BaBar, run II of the Tevatron, and the LHC can be found in Ref.~\cite{Studies}.

The outline of these lectures is as follows: in Section~\ref{sec:SM}, 
we discuss the Standard-Model description of 
$CP$ violation. After a brief look at the present status of $CP$ violation 
in the kaon system in Section~\ref{sec:kaon}, we turn to the $B$
system in Section~\ref{sec:nonlept} by giving a classification of 
non-leptonic $B$-decays and introducing low-energy effective Hamiltonians. 
A key element for $CP$ violation in the $B$ system -- the formalism
of $B^0_{d,s}$--$\overline{B^0_{d,s}}$ mixing -- is presented in 
Section~\ref{sec:mix}, and is applied to important $B$-factory benchmark 
modes in Section~\ref{sec:benchmark}. We then turn to $CP$ violation in 
charged $B$ decays in Section~\ref{sec:charged}, and discuss 
the $B_s$-meson system -- the ``El Dorado'' for hadron machines -- in 
Section~\ref{sec:Bs}. The remainder of these lectures is devoted to two 
more recent developments: the phenomenology of $B\to\pi K$ decays, which
is the topic of Section~\ref{sec:BpiK}, and the $B_d\to\pi^+\pi^-$, 
$B_s\to K^+K^-$ system, which is discussed Section~\ref{sec:Uspin}.
Before concluding in Section~\ref{sec:concl}, we make a few comments on 
other interesting rare $B$ decays in Section~\ref{sec:rare}.

\section{The Standard-Model Description of CP Violation}\label{sec:SM}
Within the Standard Model of electroweak interactions \cite{SM}, 
$CP$ violation is closely related 
to the Cabibbo--Kobayashi--Maskawa (CKM) matrix \cite{cab,km}, connecting 
the electroweak eigenstates $(d',s',b')$ of the down, strange and bottom 
quarks with their mass eigenstates $(d,s,b)$ through the following unitary 
transformation:
\begin{equation}\label{ckm}
\left(\begin{array}{c}
d'\\
s'\\
b'
\end{array}\right)=\left(\begin{array}{ccc}
V_{ud}&V_{us}&V_{ub}\\
V_{cd}&V_{cs}&V_{cb}\\
V_{td}&V_{ts}&V_{tb}
\end{array}\right)\cdot
\left(\begin{array}{c}
d\\
s\\
b
\end{array}\right)\equiv\hat V_{\mbox{{\scriptsize CKM}}}\cdot
\left(\begin{array}{c}
d\\
s\\
b
\end{array}\right).
\end{equation}
The elements of the CKM matrix describe charged-current couplings, as
can be seen easily by expressing the non-leptonic charged-current 
interaction Lagrangian in terms of the mass eigenstates appearing in
(\ref{ckm}):
\begin{equation}\label{cc-lag2}
{\cal L}_{\mbox{{\scriptsize int}}}^{\mbox{{\scriptsize CC}}}=
-\frac{g_2}{\sqrt{2}}\left(\begin{array}{ccc}\bar
u_{\mbox{{\scriptsize L}}},& \bar c_{\mbox{{\scriptsize L}}},&
\bar t_{\mbox{{\scriptsize L}}}\end{array}\right)\gamma^\mu\,\hat
V_{\mbox{{\scriptsize CKM}}}
\left(
\begin{array}{c}
d_{\mbox{{\scriptsize L}}}\\
s_{\mbox{{\scriptsize L}}}\\
b_{\mbox{{\scriptsize L}}}
\end{array}\right)W_\mu^\dagger\quad+\quad \mbox{h.c.,}
\end{equation}
where the gauge coupling $g_2$ is related to the gauge group 
$SU(2)_{\mbox{{\scriptsize L}}}$, and the $W_\mu^{(\dagger)}$ field 
corresponds to the charged $W$-bosons.

\subsection{Parametrizations of the CKM Matrix}

The phase structure of the CKM matrix is not unique, as we may perform
the following phase transformations:
\begin{equation}
V_{UD}\to\exp(i\xi_U)V_{UD}\exp(-i\xi_D),
\end{equation}
which are related to redefinitions of the up- and down-type quark fields:
\begin{equation}
U\to \exp(i\xi_U)U,\quad D\to \exp(i\xi_D)D. 
\end{equation}
Using these transformations, it can be shown that the general $N$-generation 
quark-mixing-matrix is described by 
$(N-1)^2$ parameters, which consist of $N(N-1)/2$ Euler-type angles, and
$(N-1)(N-2)/2$ complex phases. In the two-generation case \cite{cab}, 
we arrive therefore at 
\begin{equation}
\hat V_{\rm C}=\left(\begin{array}{cc}
\cos\theta_{\rm C}&\sin\theta_{\rm C}\\
-\sin\theta_{\rm C}&\cos\theta_{\rm C}
\end{array}\right),
\end{equation}
where $\sin\theta_{\rm C}=0.22$ can be determined from $K\to\pi e^+\nu_e$ 
decays.

In the case of three generations, three Euler-type angles and a single 
{\it complex phase} are needed to parametrize the CKM matrix. This complex 
phase allows us to accommodate $CP$ violation in the Standard Model, as was 
pointed out by Kobayashi and Maskawa in 1973 \cite{km}. In the 
``standard parametrization'', the three-generation CKM matrix takes the
form
\begin{equation}\label{standard}
\left(\begin{array}{ccc}
c_{12}c_{13}&s_{12}c_{13}&s_{13}e^{-i\delta_{13}}\\ -s_{12}c_{23}
-c_{12}s_{23}s_{13}e^{i\delta_{13}}&c_{12}c_{23}-
s_{12}s_{23}s_{13}e^{i\delta_{13}}&
s_{23}c_{13}\\ s_{12}s_{23}-c_{12}c_{23}s_{13}e^{i\delta_{13}}&-c_{12}s_{23}
-s_{12}c_{23}s_{13}e^{i\delta_{13}}&c_{23}c_{13}
\end{array}\right),
\end{equation}
where $c_{ij}=\cos\theta_{ij}$ and $s_{ij}=\sin\theta_{ij}$. Another
interesting parametrization of the CKM matrix was proposed by Fritzsch
and Xing \cite{FX}, which is based on the hierarchical structure of the
quark mass spectrum. 

\begin{figure}[t]
\vspace{0.10in}
\centerline{
\epsfysize=4.3truecm
\epsffile{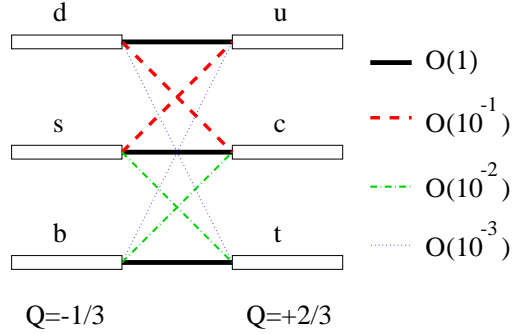}}
\caption[]{Hierarchy of the quark transitions mediated through charged 
currents.}\label{fig:term}
\end{figure}

In Fig.~\ref{fig:term}, the hierarchy of the strengths of the quark 
transitions mediated through charged-current interactions is illustrated.
In the standard parametrization (\ref{standard}), it is reflected by
\begin{equation}
s_{12}=0.22 \,\gg\, s_{23}={\cal O}(10^{-2}) \,\gg\, s_{13}={\cal O}(10^{-3}). 
\end{equation}
If we introduce new parameters $\lambda$, $A$, $\rho$ and $\eta$
by imposing the relations
\begin{equation}\label{set-rel}
s_{12}\equiv\lambda=0.22,\quad s_{23}\equiv A\lambda^2,\quad 
s_{13}e^{-i\delta}\equiv A\lambda^3(\rho-i\eta),
\end{equation}
and go back to the standard parametrization (\ref{standard}), we
arrive at
\begin{equation}
\hat V_{\mbox{{\scriptsize CKM}}} =\left(\begin{array}{ccc}
1-\frac{1}{2}\lambda^2 & \lambda & A\lambda^3(\rho-i\,\eta) \\
-\lambda & 1-\frac{1}{2}\lambda^2 & A\lambda^2\\
A\lambda^3(1-\rho-i\,\eta) & -A\lambda^2 & 1
\end{array}\right)+{\cal O}(\lambda^4).
\end{equation}
This is the ``Wolfenstein parametrization'' of the CKM matrix \cite{wolf}. 
It corresponds to an expansion in powers of the small quantity 
$\lambda=0.22$, and is very useful for phenomenological applications. 
A detailed discussion of the next-to-leading order terms in $\lambda$ 
can be found in Ref.~\cite{blo}.

\subsection{Further Requirements for CP Violation}

As we have just seen, three generations are necessary to accommodate $CP$ 
violation in the Standard Model. However, still more conditions have to be 
satisfied. They can be summarized as follows:
\begin{eqnarray}
\lefteqn{(m_t^2-m_c^2)(m_t^2-m_u^2)(m_c^2-m_u^2)}\nonumber\\
&&\times(m_b^2-m_s^2)(m_b^2-m_d^2)(m_s^2-m_d^2)\times
J_{\rm CP}\,\not=\,0,
\end{eqnarray}
where
\begin{equation}
J_{\rm CP}=\pm\,\mbox{Im}\left(V_{i\alpha}V_{j\beta}V_{i\beta}^\ast 
V_{j\alpha}^\ast\right)\quad(i\not=j,\,\alpha\not=\beta)\,.
\end{equation}
The ``Jarlskog Parameter'' $J_{\rm CP}$ represents a 
measure of the ``strength'' of $CP$ violation within the Standard Model
\cite{jarlskog}. Using the Wolfenstein parametrization, we obtain
\begin{equation}
J_{\rm CP}=\lambda^6A^2\eta={\cal O}(10^{-5}).
\end{equation}
Consequently, $CP$ violation is a small effect in the Standard Model. 
However, typically new complex couplings are present in scenarios for
new physics, yielding additional sources for $CP$ violation.

\subsection{The Unitarity Triangles of the CKM Matrix}

\begin{figure}
\begin{tabular}{lr}
   \epsfysize=3.5cm
   \epsffile{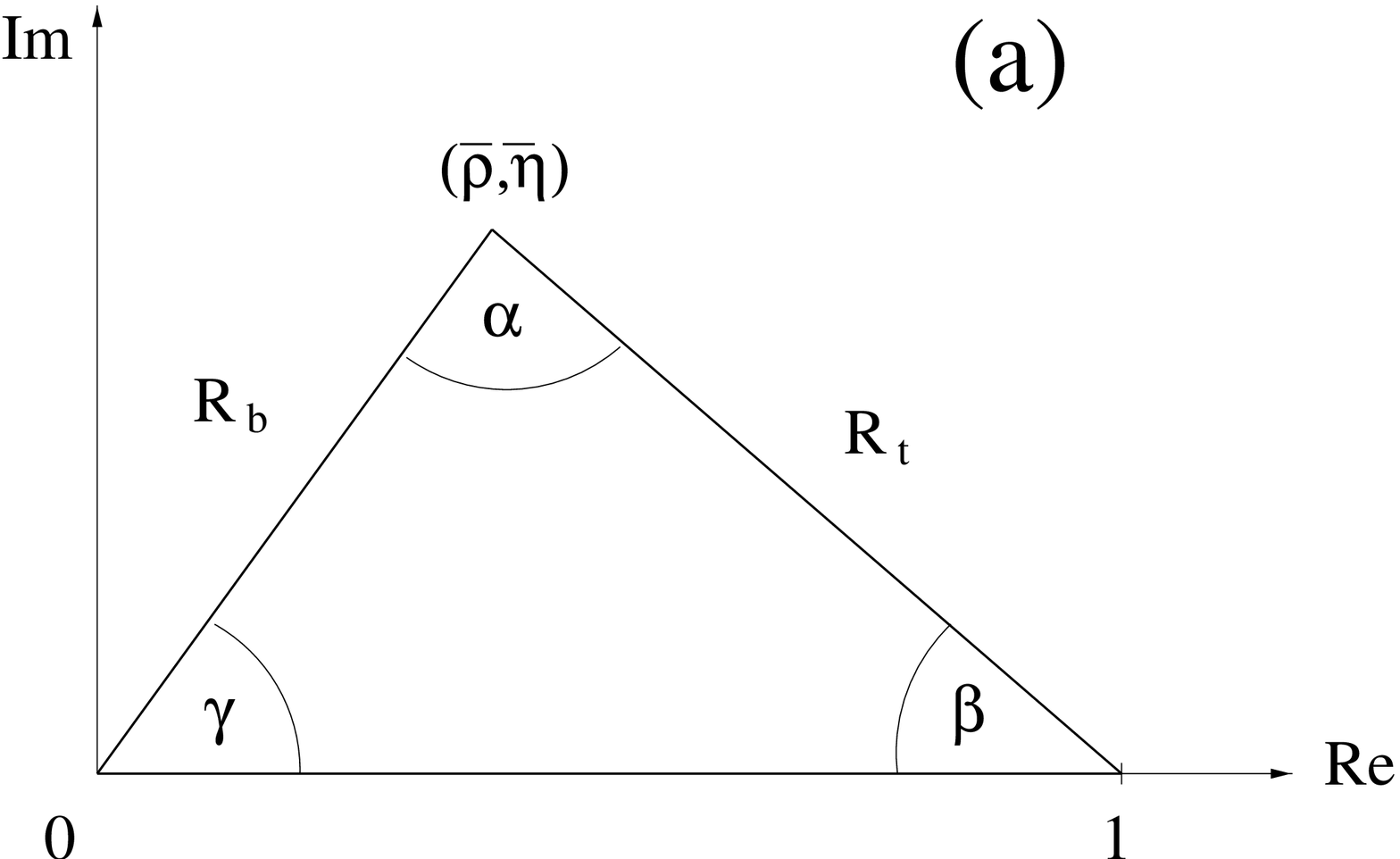}
&
   \epsfysize=3.5cm
   \epsffile{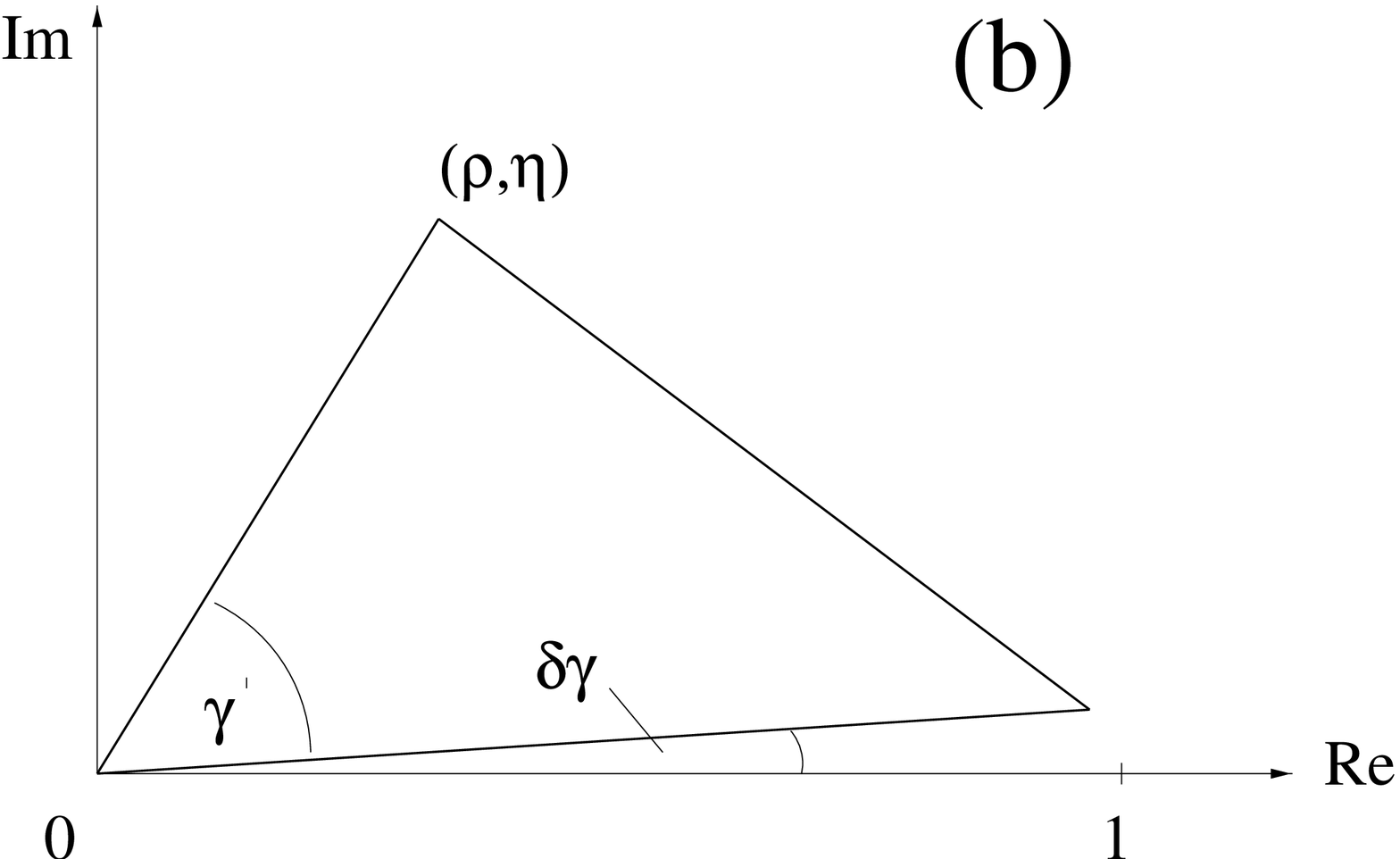}
\end{tabular}
\caption[]{The two non-squashed unitarity triangles of the CKM matrix: 
(a) and (b) correspond to the orthogonality relations (\ref{UT1}) and 
(\ref{UT2}), respectively.}
\label{fig:UT}
\end{figure}

Concerning tests of the CKM picture of $CP$ violation, the central targets 
are the {\it unitarity triangles} of the CKM matrix. The unitarity of the 
CKM matrix, which is described by
\begin{equation}
\hat V_{\mbox{{\scriptsize CKM}}}^{\,\,\dagger}\cdot\hat 
V_{\mbox{{\scriptsize CKM}}}=
\hat 1=\hat V_{\mbox{{\scriptsize CKM}}}\cdot\hat V_{\mbox{{\scriptsize 
CKM}}}^{\,\,\dagger},
\end{equation}
leads to a set of 12 equations, consisting of 6 normalization relations 
and 6 orthogonality relations. The latter can be represented as 6 triangles
in the complex plane, all having the same area, $2 A_{\Delta}=|J_{\rm CP}|$ 
\cite{AKL}. However, in only two of them, all three sides are of comparable 
magnitude ${\cal O}(\lambda^3)$, while in the remaining ones, one side is 
suppressed relative to the others by ${\cal O}(\lambda^2)$ or 
${\cal O}(\lambda^4)$. The orthogonality relations describing the 
non-squashed triangles are given by
\begin{eqnarray}
V_{ud}\,V_{ub}^\ast+V_{cd}\,V_{cb}^\ast+V_{td}\,V_{tb}^\ast&=&0
\quad\mbox{[1st and 3rd column]}\label{UT1}\\
V_{ub}^\ast\, V_{tb}+V_{us}^\ast\, V_{ts}+V_{ud}^\ast\, V_{td}&=&0
\quad\mbox{[1st and 3rd row]}.\label{UT2}
\end{eqnarray}
At leading order in $\lambda$, these relations agree with each other, and
yield
\begin{equation}\label{UTLO}
(\rho+i\eta)A\lambda^3+(-A\lambda^3)+(1-\rho-i\eta)A\lambda^3=0.
\end{equation}
Consequently, they describe the same triangle in the $\rho$--$\eta$ 
plane\footnote{Usually, the triangle relation (\ref{UTLO}) is divided by the 
overall normalization $A\lambda^3$.}, which is usually referred to 
as ``the'' unitarity triangle of the CKM matrix \cite{ut}. 
However, in the era of second-generation $B$ experiments, the experimental 
accuracy will be so tremendous that we will also have to take into account 
the next-to-leading order terms of the Wolfenstein expansion, and will have 
to distinguish between the unitarity triangles described by (\ref{UT1}) and 
(\ref{UT2}). They are illustrated in Fig.\ \ref{fig:UT}, where 
$\overline{\rho}$ and $\overline{\eta}$ are related to the Wolfenstein 
parameters $\rho$ and $\eta$ through \cite{blo}
\begin{equation}
\overline{\rho}\equiv\left(1-\lambda^2/2\right)\rho,\quad
\overline{\eta}\equiv\left(1-\lambda^2/2\right)\eta.
\end{equation}
Note that $\gamma=\gamma'+\delta\gamma$. The sides $R_b$ and $R_t$ of 
the unitarity triangle shown in Fig.\ \ref{fig:UT} (a) are given as follows:
\begin{eqnarray}
R_b&=&\left(1-\frac{\lambda^2}{2}\right)\frac{1}{\lambda}\left|
\frac{V_{ub}}{V_{cb}}\right|\,=\,\sqrt{\overline{\rho}^2+\overline{\eta}^2}
\,=\,0.41\pm0.07\label{Rb-intro}\label{Rb-def}\\
R_t&=&\frac{1}{\lambda}\left|\frac{V_{td}}{V_{cb}}\right|\,=\,
\sqrt{(1-\overline{\rho})^2+\overline{\eta}^2}\,=\,{\cal O}(1),\label{Rt-def}
\end{eqnarray}
and will show up at several places throughout these lectures.

\subsection{Towards an Allowed Range in the 
$\overline{\rho}$--$\overline{\eta}$ Plane}\label{subsec:CKM-fits}

\begin{figure}
\vspace{0.10in}
\centerline{
\epsfysize=5.0truecm
\epsffile{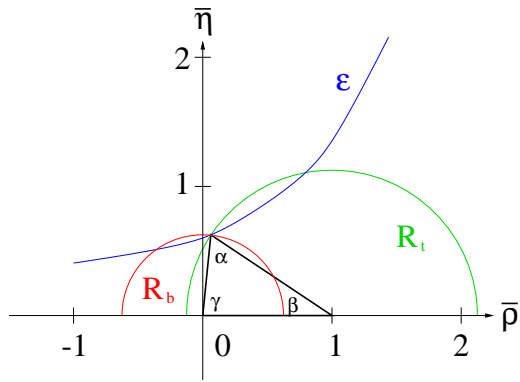}
}
\caption[]{Contours to determine the unitarity triangle in the
$\overline{\rho}$--$\overline{\eta}$ plane.}\label{fig:cont-scheme}
\end{figure}

The parameter $R_b$ introduced in (\ref{Rb-def}), i.e.\ the ratio
$|V_{ub}|/|V_{cb}|$, can be determined through semi-leptonic $b\to u$
and $b\to c$ decays. It fixes a circle in the 
$\overline{\rho}$--$\overline{\eta}$ plane around $(0,0)$ with radius 
$R_b$. The second side $R_t$ of the unitarity triangle shown in 
Fig.\ \ref{fig:UT} (a) can be determined through 
$B^0_{d,s}$--$\overline{B^0_{d,s}}$ mixing. It fixes another
circle in the $\overline{\rho}$--$\overline{\eta}$ plane, which is
centered at $(1,0)$ and has radius $R_t$. Finally, using experimental 
information on an observable $\varepsilon$, which describes ``indirect'' 
$CP$ violation in the neutral kaon system and will be discussed in the 
next section, a hyperbola in the $\overline{\rho}$--$\overline{\eta}$ 
plane can be fixed. These contours are sketched in 
Fig.~\ref{fig:cont-scheme}; their intersection gives the 
apex of the unitarity triangle shown in Fig.\ \ref{fig:UT} (a). The
contours that are implied by $B^0_{d}$--$\overline{B^0_{d}}$ mixing and
$\varepsilon$ depend on $|V_{cb}|$, the top-quark mass, QCD corrections,
and non-perturbative parameters (for a review, see \cite{BF-rev}). This 
feature leads to strong correlations between theoretical and experimental 
uncertainties. A detailed recent analysis was performed by Ali and London 
\cite{AL}, who find the following ranges: 
\begin{equation}\label{UT-range}
75^\circ\leq\alpha\leq121^\circ,\quad
16^\circ\leq\beta\leq 34^\circ, \quad
38^\circ\leq\gamma\leq81^\circ.
\end{equation}
We shall come back to this issue in Subsection~\ref{Bs-gen}, where we 
emphasize that the present experimental lower bound on 
$B^0_{s}$--$\overline{B^0_{s}}$ mixing has already a very important 
impact on the allowed range in the $\overline{\rho}$--$\overline{\eta}$ 
plane (see Fig.~\ref{fig:UT-constr}).

\section{A Brief Look at CP Violation in the Kaon System}\label{sec:kaon}
Although the discovery of $CP$ violation goes back to 1964 \cite{CP-discovery},
so far this phenomenon could only be observed in the $K$-meson system. Here 
it is described by two complex quantities, called $\varepsilon$ and 
$\varepsilon'$, which are defined by the following ratios of decay 
amplitudes:
\begin{equation}\label{defs-eps}
\frac{A(K_{\rm L}\to\pi^+\pi^-)}{A(K_{\rm S}
\to\pi^+\pi^-)}=\varepsilon+\varepsilon',\quad
\frac{A(K_{\rm L}\to\pi^0\pi^0)}{A(K_{\rm S}
\to\pi^0\pi^0)}=\varepsilon-2\,\varepsilon'.
\end{equation}
While $\varepsilon=(2.280\pm0.013)\times e^{i\frac{\pi}{4}}\times 10^{-3}$
parametrizes ``indirect'' $CP$ violation, originating from the fact that
the mass eigenstates of the neutral kaon system are not $CP$ eigenstates,
the quantity Re$(\varepsilon'/\varepsilon)$ measures ``direct'' $CP$ violation 
in $K\to\pi\pi$ transitions. The $CP$-violating observable $\varepsilon$ plays 
an important role to constrain the unitarity triangle \cite{BF-rev,AL} and 
implies  -- using reasonable assumptions about certain hadronic parameters --
in particular a positive value of the Wolfenstein parameter $\eta$. In 1999,
new measurements of Re$(\varepsilon'/\varepsilon)$ have demonstrated that
this observable is non-zero, thereby excluding ``superweak''
models of $CP$ violation \cite{superweak}:
\begin{equation}\label{epsprime-res}
\mbox{Re}(\varepsilon'/\varepsilon)=\left\{\begin{array}{ll}
(28\pm4.1)\times10^{-4}&\mbox{(KTeV Collaboration \cite{KTeV}),}\\
(14\pm4.3)\times10^{-4}&\mbox{(NA48 Collaboration \,\cite{NA48}).}
\end{array}\right.
\end{equation}
Unfortunately, the calculations of Re$(\varepsilon'/\varepsilon)$ are very 
involved and suffer at present from large hadronic uncertainties 
\cite{epsprime}. Consequently, this observable does not allow a powerful 
test of the $CP$-violating sector of the Standard Model, unless the hadronic 
matrix elements of the relevant operators can be brought under better control. 

In order to test the Standard-Model description of $CP$ violation, the 
rare decays $K_{\rm L}\to\pi^0\nu\overline{\nu}$ and 
$K^+\to\pi^+\nu\overline{\nu}$ are more promising, and may allow a 
determination of $\sin(2\beta)$ with respectable accuracy \cite{bb}.
Yet it is clear that the kaon system by itself cannot provide the whole 
picture of $CP$ violation, and therefore it is essential to study $CP$ 
violation outside this system. In this respect, $B$-meson decays appear 
to be most promising. There are of course also other interesting probes
to explore $CP$ violation, for example, the neutral $D$-meson system
or electric dipole moments, which will, however, not be addressed further
in these lectures.

\section{Non-leptonic B Decays}\label{sec:nonlept}
With respect to testing the Standard-Model description of $CP$ violation, 
the major role is played by non-leptonic $\overline{B}$ decays, which are 
mediated by $b\to q_1\,\overline{q_2}\,d\,(s)$ quark-level 
transitions ($q_1,q_2\in\{u,d,c,s\}$).

\subsection{Classification}\label{sec:class}

\begin{figure}[ht]
\begin{center}
{\small
\begin{picture}(140,60)(0,20)
\Line(10,50)(130,50)\Text(5,50)[r]{$b$}\Text(140,50)[l]{$d\,(s)$}
\PhotonArc(70,50)(30,0,180){3}{15}
\Text(69,56)[b]{$u,c,t$}\Text(109,75)[b]{$W$}
\Gluon(70,50)(120,10){2}{10}
\Line(120,10)(135,23)\Line(120,10)(135,-3)
\Text(85,22)[tr]{$G$}\Text(140,-3)[l]{$q_1$}
\Text(140,23)[l]{$\overline{q_2}=\overline{q_1}$}
\end{picture}}
\end{center}
\vspace*{0.7truecm}
\caption[]{QCD penguin diagrams ($q_1=q_2\in\{u,d,c,s\}$).}\label{fig:QCD-top}
\end{figure}
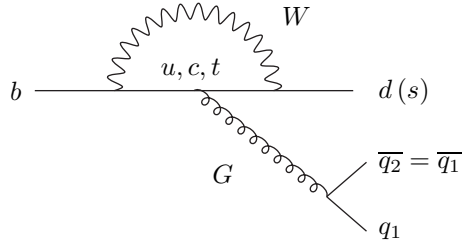

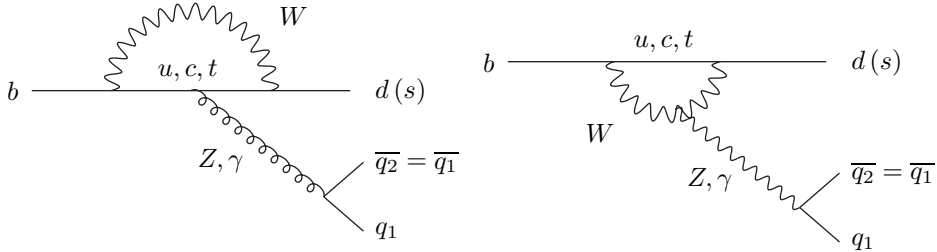
\begin{figure}
{\small
\begin{picture}(140,60)(0,20)
\Line(10,50)(130,50)\Text(5,50)[r]{$b$}\Text(140,50)[l]{$d\,(s)$}
\PhotonArc(70,50)(30,0,180){3}{15}
\Text(69,56)[b]{$u,c,t$}\Text(109,75)[b]{$W$}
\Gluon(70,50)(120,10){2}{10}
\Line(120,10)(135,23)\Line(120,10)(135,-3)
\Text(90,28)[tr]{$Z,\gamma$}\Text(140,-3)[l]{$q_1$}
\Text(140,23)[l]{$\overline{q_2}=\overline{q_1}$}
\end{picture}}

\vspace*{-2truecm}\hspace*{6.2truecm}
{\small
\begin{picture}(140,60)(0,20)
\Line(10,65)(130,65)\Text(5,65)[r]{$b$}\Text(140,65)[l]{$d\,(s)$}
\PhotonArc(70,65)(20,180,360){3}{10}
\Text(69,71)[b]{$u,c,t$}\Text(45,35)[b]{$W$}
\Photon(73,47)(120,10){2}{10}
\Line(120,10)(135,23)\Line(120,10)(135,-3)
\Text(95,25)[tr]{$Z,\gamma$}\Text(140,-3)[l]{$q_1$}
\Text(140,23)[l]{$\overline{q_2}=\overline{q_1}$}
\end{picture}}
\vspace*{0.7truecm}
\caption[]{Electroweak penguin diagrams 
($q_1=q_2\in\{u,d,c,s\}$).}\label{fig:EWP-top}
\end{figure}

There are two kinds of topologies contributing to non-leptonic $B$ decays: 
tree-diagram-like and ``penguin'' topologies. The latter consist of gluonic 
(QCD) and electroweak (EW) penguins. In 
Figs.~\ref{fig:tree-top}--\ref{fig:EWP-top}, the corresponding 
leading-order Feynman diagrams are shown. Depending
on the flavour content of their final states, we may classify 
$b\to q_1\,\overline{q_2}\,d\,(s)$ decays as follows:
\begin{itemize}
\item $q_1\not=q_2\in\{u,c\}$: only tree diagrams contribute.
\item $q_1=q_2\in\{u,c\}$: tree and penguin diagrams contribute.
\item $q_1=q_2\in\{d,s\}$: only penguin diagrams contribute.
\end{itemize}

\begin{figure}[ht]
\begin{center}
{\small
\vspace*{-1truecm}\hspace*{4truecm}\begin{picture}(80,50)(80,20)
\Line(10,45)(80,45)\Photon(40,45)(80,5){2}{10}
\Line(80,5)(105,20)\Line(80,5)(105,-10)
\Text(5,45)[r]{$b$}\Text(85,45)[l]{$q_1$}
\Text(110,20)[l]{$\overline{q_2}$}
\Text(110,-10)[l]{$d\,(s)$}
\Text(45,22)[tr]{$W$}
\end{picture}}
\end{center}
\vspace*{1.0truecm}
\caption[]{Tree diagrams ($q_1,q_2\in\{u,c\}$).}\label{fig:tree-top}
\end{figure}
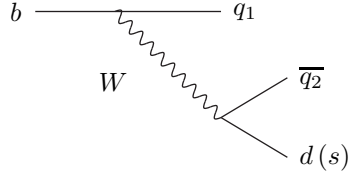

\subsection{Low-energy Effective Hamiltonians}\label{subsec:ham}
In order to analyse non-leptonic $B$ decays theoretically, one uses 
low-energy effective Hamiltonians, which are calculated by making use 
of the operator product expansion, yielding transition matrix elements 
of the following structure:
\begin{equation}\label{ee2}
\langle f|{\cal H}_{\rm eff}|i\rangle=\frac{G_{\rm F}}{\sqrt{2}}
\lambda_{\rm CKM}\sum_k C_k(\mu)\langle f|Q_k(\mu)|i\rangle\,.
\end{equation}
The operator product expansion allows us to separate the short-distance
contributions to this transition amplitude from the long-distance 
ones, which are described by perturbative Wilson coefficient 
functions $C_k(\mu)$ and non-perturbative hadronic matrix elements 
$\langle f|Q_k(\mu)|i\rangle$, respectively. As usual, $G_{\rm F}$ is
the Fermi constant, $\lambda_{\rm CKM}$ is a CKM factor, and $\mu$ denotes an 
appropriate renormalization scale. The $Q_k$ are local operators, which 
are generated by QCD and electroweak interactions and govern ``effectively'' 
the decay in question. The Wilson coefficients $C_k(\mu)$ can be 
considered as scale-dependent couplings related to the vertices described
by the $Q_k$.

\begin{figure}
\vspace*{-0.4truecm}
\begin{center}
\leavevmode
\epsfysize=4.0truecm 
\epsffile{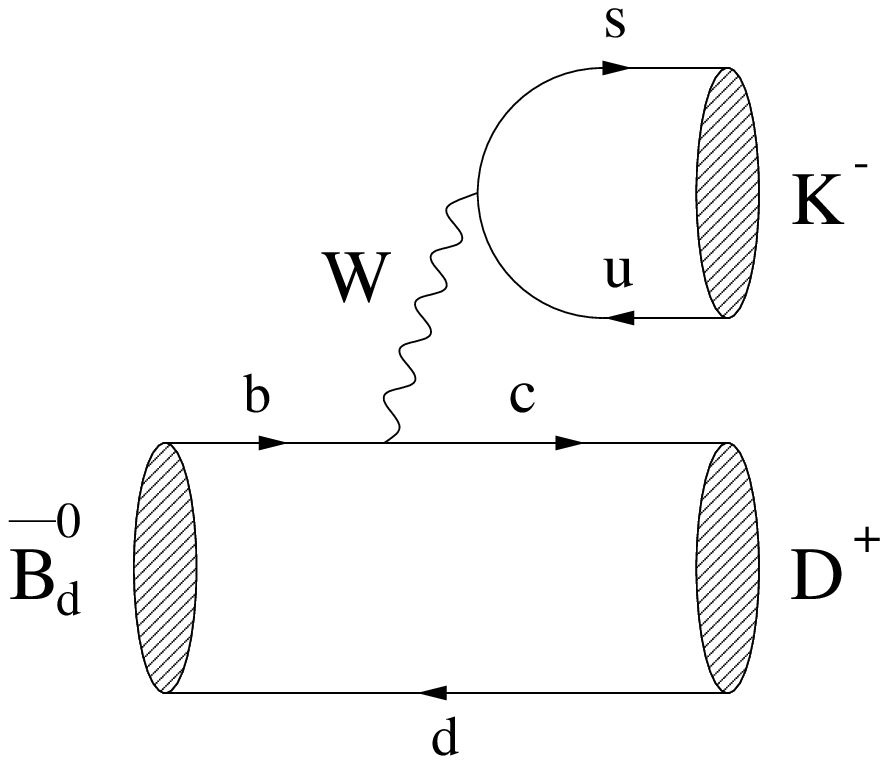} \hspace*{2truecm}
\epsfysize=3.5truecm 
\epsffile{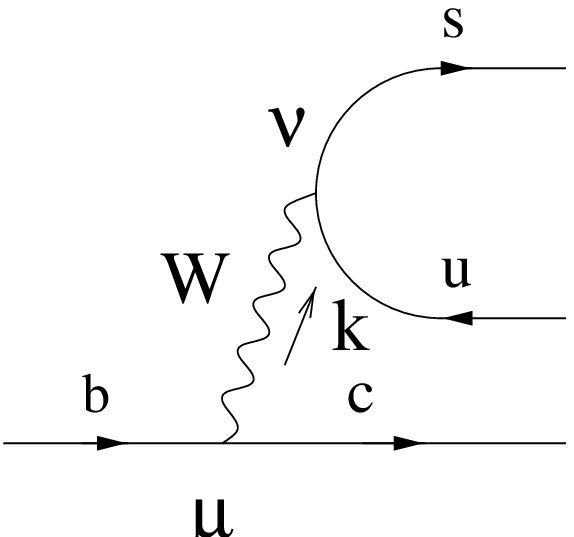}
\vspace*{-0.3truecm}
\caption{Feynman diagram contributing at leading order to 
$\overline{B^0_d}\to D^+K^-$.}\label{fig:feyn-example}
\end{center}
\end{figure}

\begin{figure}[b]
\vspace*{-0.6truecm}
\begin{center}
\leavevmode
\epsfysize=2.8truecm 
\epsffile{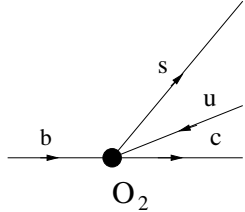}
\vspace*{-0.3truecm}
\caption{The vertex described by the ``current--current'' operator 
$O_2$.}\label{fig:feyn-example2}
\end{center}
\end{figure}

Let us consider $\overline{B^0_d}\to D^+K^-$, a pure ``tree'' decay, 
as an example. At leading order, this transition originates from the 
process shown in Fig.~\ref{fig:feyn-example}. Evaluating the corresponding 
Feynman diagram yields the amplitude
\begin{equation}\label{trans-ampl}
-\,\frac{g_2^2}{8}V_{us}^\ast V_{cb}
\left[\overline{s}\gamma^\nu(1-\gamma_5)u\right]
\left[\frac{g_{\nu\mu}}{k^2-M_W^2}\right]
\left[\overline{c}\gamma^\mu(1-\gamma_5)b\right].
\end{equation}
As $k^2\approx m_b^2\ll M_W^2$, we have 
\begin{equation}
\frac{g_{\nu\mu}}{k^2-M_W^2}\quad\longrightarrow\quad
-\,\frac{g_{\nu\mu}}{M_W^2}\equiv-\left(\frac{8G_{\rm F}}{\sqrt{2}g_2^2}
\right)g_{\nu\mu},
\end{equation}
i.e.\ we may ``integrate out'' the $W$-boson in (\ref{trans-ampl}), 
and arrive at
\begin{eqnarray}
\lefteqn{{\cal H}_{\rm eff}=\frac{G_{\rm F}}{\sqrt{2}}V_{us}^\ast V_{cb}
\left[\overline{s}_\alpha\gamma_\mu(1-\gamma_5)u_\alpha\right]
\left[\overline{c}_\beta\gamma^\mu(1-\gamma_5)b_\beta\right]}\nonumber\\
&&=\frac{G_{\rm F}}{\sqrt{2}}V_{us}^\ast V_{cb}
(\overline{s}_\alpha u_\alpha)_{\mbox{{\scriptsize 
V--A}}}(\overline{c}_\beta b_\beta)_{\mbox{{\scriptsize V--A}}}
\equiv\frac{G_{\rm F}}{\sqrt{2}}V_{us}^\ast V_{cb}O_2\,,
\end{eqnarray}
where $\alpha$ and $\beta$ denote $SU(3)_{\rm C}$ colour indices. 
Effectively, the vertex shown in Fig.~\ref{fig:feyn-example} is now
described by the ``current--current'' operator $O_2$ (see 
Fig.~\ref{fig:feyn-example2}).

If we take into account QCD corrections, operator mixing leads to a second
``current--current'' operator:
\begin{equation}
O_1\equiv\left[\overline{s}_\alpha\gamma_\mu(1-\gamma_5)u_\beta\right]
\left[\overline{c}_\beta\gamma^\mu(1-\gamma_5)b_\alpha\right].
\end{equation}
Consequently, we obtain a low-energy effective Hamiltonian of the following
structure:
\begin{equation}\label{Heff-example}
{\cal H}_{\rm eff}=\frac{G_{\rm F}}{\sqrt{2}}V_{us}^\ast V_{cb}
\left[C_1(\mu)O_1+C_2(\mu)O_2\right],
\end{equation}
where $C_1(\mu)\not=0$ and $C_2(\mu)\not=1$ are due to QCD renormalization
effects. In order to evaluate these coefficients, we first have to 
calculate QCD corrections to the vertices shown in 
Figs.~\ref{fig:feyn-example} and \ref{fig:feyn-example2}, and then have to 
express the QCD-corrected transition amplitude in terms of QCD-corrected
matrix elements and Wilson coefficients as in (\ref{ee2}). This procedure
is called ``matching''. The results for the $C_k(\mu)$ obtained this way 
contain terms of $\mbox{log}(\mu/M_W)$, which become large for 
$\mu={\cal O}(m_b)$, the scale governing the hadronic matrix elements
of the $O_k$. However, the renormalization group, exploiting the fact that
the transition amplitude (\ref{ee2}) cannot depend on the chosen 
renormalization scale $\mu$, allows us to sum up the following terms
of the Wilson coefficients:
\begin{equation}
\alpha_s^n\left[\log\left(\frac{\mu}{M_W}\right)\right]^n 
\,\,\mbox{(LO)},\quad\,\,\alpha_s^n\left[\log\left(\frac{\mu}{M_W}\right)
\right]^{n-1}\,\,\mbox{(NLO)},\quad ...
\end{equation}
A detailed discussion of these techniques can be found in 
Refs.~\cite{Buras-lect,BBL-rev}.

\subsection{Decays with Tree and Penguin Contributions}
In order to explore $CP$ violation in the $B$ system, non-leptonic
decays receiving both tree and penguin contributions, i.e.\
$|\Delta B|=1$, $\Delta C=\Delta U=0$ modes (see Subsection~\ref{sec:class}),
play an outstanding role. In this case, because of the penguin topologies,
the operator basis is much larger than in our example 
(\ref{Heff-example}), where we considered a pure ``tree'' decay. We obtain
\begin{equation}\label{e4}
{\cal H}_{\mbox{{\scriptsize eff}}}=\frac{G_{\mbox{{\scriptsize 
F}}}}{\sqrt{2}}\left[\sum\limits_{j=u,c}V_{jr}^\ast V_{jb}\left\{\sum
\limits_{k=1}^2C_k(\mu)\,Q_k^{jr}+\sum\limits_{k=3}^{10}C_k(\mu)\,Q_k^{r}
\right\}\right],
\end{equation}
where the operators $Q_k^{jr}$ ($j\in\{u,c\}$, $r\in\{d,s\}$) can 
be divided as follows:
\begin{itemize}
\item Current--current operators:
\begin{equation}
\begin{array}{rcl}
Q_{1}^{jr}&=&(\overline{r}_{\alpha}j_{\beta})_{\mbox{{\scriptsize V--A}}}
(\overline{j}_{\beta}b_{\alpha})_{\mbox{{\scriptsize V--A}}}\\
Q_{2}^{jr}&=&(\overline{r}_\alpha j_\alpha)_{\mbox{{\scriptsize 
V--A}}}(\overline{j}_\beta b_\beta)_{\mbox{{\scriptsize V--A}}}.
\end{array}
\end{equation}
\item QCD penguin operators:
\begin{equation}\label{qcd-penguins}
\begin{array}{rcl}
Q_{3}^r&=&(\overline{r}_\alpha b_\alpha)_{\mbox{{\scriptsize V--A}}}\sum_{q'}
(\overline{q}'_\beta q'_\beta)_{\mbox{{\scriptsize V--A}}}\\
Q_{4}^r&=&(\overline{r}_{\alpha}b_{\beta})_{\mbox{{\scriptsize V--A}}}
\sum_{q'}(\overline{q}'_{\beta}q'_{\alpha})_{\mbox{{\scriptsize V--A}}}\\
Q_{5}^r&=&(\overline{r}_\alpha b_\alpha)_{\mbox{{\scriptsize V--A}}}\sum_{q'}
(\overline{q}'_\beta q'_\beta)_{\mbox{{\scriptsize V+A}}}\\
Q_{6}^r&=&(\overline{r}_{\alpha}b_{\beta})_{\mbox{{\scriptsize V--A}}}
\sum_{q'}(\overline{q}'_{\beta}q'_{\alpha})_{\mbox{{\scriptsize V+A}}}.
\end{array}
\end{equation}
\item Electroweak (EW) penguin operators (the $e_{q'}$ denote quark charges):
\begin{equation}
\begin{array}{rcl}
Q_{7}^r&=&\frac{3}{2}(\overline{r}_\alpha b_\alpha)_{\mbox{{\scriptsize V--A}}}
\sum_{q'}e_{q'}(\overline{q}'_\beta q'_\beta)_{\mbox{{\scriptsize V+A}}}\\
Q_{8}^r&=&
\frac{3}{2}(\overline{r}_{\alpha}b_{\beta})_{\mbox{{\scriptsize V--A}}}
\sum_{q'}e_{q'}(\overline{q}_{\beta}'q'_{\alpha})_{\mbox{{\scriptsize V+A}}}\\
Q_{9}^r&=&\frac{3}{2}(\overline{r}_\alpha b_\alpha)_{\mbox{{\scriptsize V--A}}}
\sum_{q'}e_{q'}(\overline{q}'_\beta q'_\beta)_{\mbox{{\scriptsize V--A}}}\\
Q_{10}^r&=&
\frac{3}{2}(\overline{r}_{\alpha}b_{\beta})_{\mbox{{\scriptsize V--A}}}
\sum_{q'}e_{q'}(\overline{q}'_{\beta}q'_{\alpha})_{\mbox{{\scriptsize V--A}}}.
\end{array}
\end{equation}
\end{itemize}
The current--current, QCD and EW penguin operators are related to the tree, 
QCD and EW penguin processes shown in 
Figs.~\ref{fig:tree-top}--\ref{fig:EWP-top}. At a renormalization scale
$\mu={\cal O}(m_b)$, the Wilson coefficients of the current--current operators
are $C_1(\mu)={\cal O}(10^{-1})$ and $C_2(\mu)={\cal O}(1)$, whereas those
of the penguin operators are ${\cal O}(10^{-2})$. The calculation of 
(\ref{e4}) beyond the leading logarithmic approximation (LO) has been 
reviewed in \cite{BBL-rev}, where also numerical values of the 
corresponding (NLO) Wilson coefficients are given.

\subsection{Electroweak Penguin Effects}
Since the ratio $\alpha/\alpha_s={\cal O}(10^{-2})$ of the QED and QCD 
couplings is very small, we expect naively that EW penguins should play a 
minor role in comparison with QCD penguins. This would actually be the 
case if the top quark was not ``heavy''. However, since the Wilson 
coefficient of the EW penguin operator $Q_9$ increases strongly with the 
top-quark mass $m_t$, we obtain interesting EW penguin effects in several
$B$ decays: $B^-\to K^-\phi$ is affected significantly by EW
penguins, whereas $B\to\pi\phi$ and $B_s\to\pi^0\phi$ are even dominated 
by such topologies \cite{RF-EWP,RF-rev}. Electroweak penguins have also
an important impact on $B\to\pi K$ modes \cite{EWP-BpiK}, as we will 
discuss in more detail in Section~\ref{sec:BpiK}. 

\subsection{Factorization of Hadronic Matrix Elements}
In order to discuss ``factorization'', let us consider again our example
from Subsection~\ref{subsec:ham}, the decay $\overline{B^0_d}\to D^+K^-$.
The problem in the evaluation of the corresponding transition amplitude
is the calculation of the hadronic matrix elements of the $O_{1,2}$ operators
between the $\langle K^-D^+|$ final and $|\overline{B^0_d}\rangle$ initial 
states. Making use of the well-known $SU(N_{\rm C})$ colour-algebra relation
\begin{equation}
T^a_{\alpha\beta}T^a_{\gamma\delta}=\frac{1}{2}\left(\delta_{\alpha\delta}
\delta_{\beta\gamma}-\frac{1}{N_{\rm C}}\delta_{\alpha\beta}
\delta_{\gamma\delta}\right)
\end{equation}
to re-write the operator $O_1$, we obtain
\begin{displaymath}
\langle K^-D^+|{\cal H}_{\rm eff}|\overline{B^0_d}\rangle=
\frac{G_{\rm F}}{\sqrt{2}}V_{us}^\ast V_{cb}\Bigl[a_1\langle K^-D^+|
(\overline{s}_\alpha u_\alpha)_{\mbox{{\scriptsize V--A}}}
(\overline{c}_\beta b_\beta)_{\mbox{{\scriptsize V--A}}}
|\overline{B^0_d}\rangle
\end{displaymath}
\vspace*{-0.3truecm}
\begin{equation}
+2\,C_1\langle K^-D^+|
(\overline{s}_\alpha\, T^a_{\alpha\beta}\,u_\beta)_{\mbox{{\scriptsize 
V--A}}}(\overline{c}_\gamma 
\,T^a_{\gamma\delta}\,b_\delta)_{\mbox{{\scriptsize V--A}}}
|\overline{B^0_d}\rangle\Bigr],\nonumber
\end{equation}
with
\begin{equation}\label{a1-def}
a_1=\frac{C_1}{N_{\rm C}}+C_2.
\end{equation}
It is now straightforward to ``factorize'' the hadronic matrix elements:
\begin{eqnarray}
\lefteqn{\left.\langle K^-D^+|
(\overline{s}_\alpha u_\alpha)_{\mbox{{\scriptsize 
V--A}}}(\overline{c}_\beta b_\beta)_{\mbox{{\scriptsize V--A}}}
|\overline{B^0_d}\rangle\right|_{\rm fact}}\nonumber\\
&&=\langle K^-|\left[\overline{s}_\alpha\gamma_\mu(1-\gamma_5)u_\alpha\right]
|0\rangle\langle D^+|\left[\overline{c}_\beta\gamma^\mu
(1-\gamma_5)b_\beta\right]|\overline{B^0_d}\rangle\nonumber\\
&&\propto f_K \mbox{(``decay constant'')}\times 
F_{BD} \mbox{(``form factor'')},
\end{eqnarray}
\begin{equation}
\left.\langle K^-D^+|
(\overline{s}_\alpha\, T^a_{\alpha\beta}\,u_\beta)_{\mbox{{\scriptsize 
V--A}}}(\overline{c}_\gamma 
\,T^a_{\gamma\delta}\,b_\delta)_{\mbox{{\scriptsize V--A}}}
|\overline{B^0_d}\rangle\right|_{\rm fact}=0.
\end{equation}
The quantity introduced in (\ref{a1-def}) is a phenomenological
``colour factor'', governing ``colour-allowed'' decays. In the case of
``colour-suppressed'' modes, for instance $\overline{B^0_d}\to
\pi^0D^0$, we have to deal with the combination
\begin{equation}\label{a2-def}
a_2=C_1+\frac{C_2}{N_{\rm C}}.
\end{equation}

The concept of ``factorization'' of hadronic matrix elements has
a long history \cite{facto}, and can be justified, for example, 
in the large $N_{\rm C}$ limit \cite{largeN}. Recently, an interesting
approach was proposed in Ref.~\cite{BBNS}, which may provide an important 
step towards a rigorous basis for factorization for a large class of 
non-leptonic two-body $B$-meson decays in the heavy-quark limit. The 
resulting ``factorization'' formula incorporates elements both of the 
``naive'' factorization approach sketched above and of the hard-scattering 
picture. Let us consider a decay $\overline{B}\to M_1M_2$, where $M_1$ picks 
up the spectator quark. If $M_1$ is either a heavy ($D$) or a light 
($\pi$, $K$) meson, and $M_2$ a light ($\pi$, $K$) meson,
a ``QCD factorization'' formula of the following structure can be 
derived:\footnote{``QCD factorization'' does not hold, if $M_2$ is a 
heavy ($D$) meson.}
\begin{displaymath}
A(\overline{B}\to M_1M_2)=\left[\mbox{``naive factorization''}\right]
\end{displaymath}
\begin{equation}\label{QCD-factor}
~~~~~~~~\times
\left[1\,+\,\mbox{calculable ${\cal O}(\alpha_s)$}\,+\,
{\cal O}(\Lambda_{\rm QCD}/m_b)\right].
\end{equation}
Whereas the ${\cal O}(\alpha_s)$ terms, i.e.\ the radiative 
non-factorizable corrections to ``naive'' factorization, can be 
calculated in a systematic way, the main limitation is due to the 
${\cal O}(\Lambda_{\rm QCD}/m_b)$ terms, which require further studies.

\begin{figure}
\begin{center}
{\small
\begin{picture}(250,70)(0,45)
\ArrowLine(10,100)(40,100)\Photon(40,100)(80,100){2}{8}
\ArrowLine(80,100)(110,100)
\ArrowLine(40,60)(10,60)\Photon(40,60)(80,60){2}{8}
\ArrowLine(110,60)(80,60)
\ArrowLine(40,100)(40,60)\ArrowLine(80,60)(80,100)
\Text(25,105)[b]{$q$}\Text(60,105)[b]{$W$}\Text(95,105)[b]{$b$}
\Text(25,55)[t]{$b$}\Text(60,55)[t]{$W$}\Text(95,55)[t]{$q$}
\Text(35,80)[r]{$u,c,t$}\Text(85,80)[l]{$u,c,t$}
\ArrowLine(160,100)(190,100)\ArrowLine(190,100)(230,100)
\ArrowLine(230,100)(260,100)
\ArrowLine(190,60)(160,60)\ArrowLine(230,60)(190,60)
\ArrowLine(260,60)(230,60)
\Photon(190,100)(190,60){2}{8}\Photon(230,60)(230,100){2}{8}
\Text(175,105)[b]{$q$}\Text(245,105)[b]{$b$}
\Text(175,55)[t]{$b$}\Text(245,55)[t]{$q$}
\Text(210,105)[b]{$u,c,t$}\Text(210,55)[t]{$u,c,t$}
\Text(180,80)[r]{$W$}\Text(240,80)[l]{$W$}
\end{picture}}
\end{center}
\vspace*{-0.4truecm}
\caption{Box diagrams contributing to $B^0_q$--$\overline{B^0_q}$ mixing 
($q\in\{d,s\}$).}\label{fig:boxes}
\end{figure}
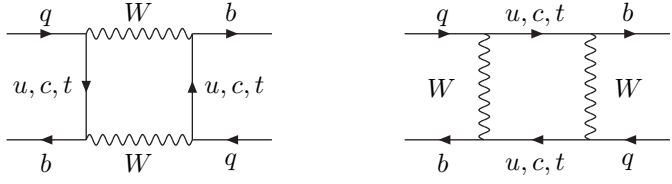

\boldmath
\section{The Formalism of $B^0_{d,s}$--$\overline{B^0_{d,s}}$ 
Mixing}\label{sec:mix}\unboldmath
Within the Standard Model, $B^0_q$--$\overline{B^0_q}$ mixing 
($q\in\{d,s\}$) is induced at lowest order through the box diagrams 
shown in Fig.~\ref{fig:boxes}. The Wigner--Weisskopf formalism yields 
an effective Schr{\"o}dinger equation
\begin{equation}\label{SG-OSZ}
i\,\frac{\partial}{\partial t}\left(\begin{array}{c} a(t)\\ b(t)
\end{array}
\right)=\left[\left(\begin{array}{cc}
M_{0}^{(q)} & M_{12}^{(q)}\\ M_{12}^{(q)\ast} & M_{0}^{(q)}
\end{array}\right)-
\frac{i}{2}\left(\begin{array}{cc}
\Gamma_{0}^{(q)} & \Gamma_{12}^{(q)}\\
\Gamma_{12}^{(q)\ast} & \Gamma_{0}^{(q)}
\end{array}\right)\right]
\cdot\left(\begin{array}{c}
a(t)\\ b(t)\nonumber
\end{array}
\right),
\end{equation}
which describes the time evolution of the state vector
\begin{equation}
\left\vert \psi_q(t)\right\rangle=a(t)\left\vert B^{0}_q\right\rangle+
b(t)\left\vert\overline{B^{0}_q}\right\rangle.
\end{equation}

\subsection{Solution of the Schr\"odinger Equation}
It is straightforward to calculate the eigenstates 
$\left\vert B_{\pm}^{(q)}\right\rangle$ and eigenvalues 
$\lambda_{\pm}^{(q)}$ of the Hamiltonian given in (\ref{SG-OSZ}):
\begin{equation}
\left\vert B_{\pm}^{(q)} \right\rangle  =
\frac{1}{\sqrt{1+\vert \alpha_q\vert^{2}}}
\left(\left\vert B^{0}_q\right\rangle\pm\alpha_q\left\vert
\overline{B^{0}_q}\right\rangle\right)
\end{equation}
\begin{equation}
\lambda_{\pm}^{(q)}  =
\left(M_{0}^{(q)}-\frac{i}{2}\Gamma_{0}^{(q)}
\right)\pm
\left(M_{12}^{(q)}-\frac{i}{2}\Gamma_{12}^{(q)}\right)\alpha_q,
\end{equation}
where
\begin{equation}
\alpha_q e^{+i\left(\Theta_{\Gamma_{12}}^{(q)}+n'
\pi\right)}=
\sqrt{\frac{4\left\vert M_{12}^{(q)}\right\vert^{2}
e^{-i2\delta\Theta_{M/\Gamma}^{(q)}}+\left\vert\Gamma_{12}^{(q)}\right
\vert^{2}}{4\left\vert M_{12}^{(q)}\right\vert^{2}+\left\vert
\Gamma_{12}^{(q)}\right\vert^{2}- 4\left\vert M_{12}^{(q)}\right\vert
\left\vert\Gamma_{12}^{(q)}\right\vert\sin\delta\Theta_{M
/\Gamma}^{(q)}}}.
\end{equation}
Here $M_{12}^{(q)}\equiv e^{i\Theta_{M_{12}}^{(q)}}\vert
M_{12}^{(q)}\vert$, $\Gamma_{12}^{(q)}\equiv
e^{i\Theta_{\Gamma_{12}}^{(q)}}\vert\Gamma_{12}^{(q)}
\vert$ and $\delta\Theta_{M/\Gamma}^{(q)}\equiv\Theta_{M_{12}}^{(q)}-
\Theta_{\Gamma_{12}}^{(q)}$. The $n'\in\{0,1\}$ parametrizes the sign 
of the square root appearing in that expression. Calculating the dispersive 
parts of the box diagrams gives \cite{BF-rev}
\begin{equation}
M_{12}^{(q)}=
\frac{G_{\mbox{{\scriptsize F}}}^{2}M_{W}^{2}\eta_B m_{B_q}
B_{B_q}F_{B_q}^{2}}{12\pi^{2}}\left(V_{tq}^\ast V_{tb}\right)^2
S_0(x_{t})\,e^{i(\pi-\phi_{\mbox{{\scriptsize CP}}}(B_q))},
\end{equation}
where $\eta_B$ is a perturbative QCD correction, $m_{B_q}$ the $B_q$-meson 
mass, $B_{B_q}$ a non-perturbative ``bag'' parameter related to 
$\langle\overline{B^0_q}|[\overline{b}\,\gamma_\mu(1-\gamma_5)q]^2
|B^0_q\rangle$, $F_{B_q}$ the $B_q$ decay constant,  
$x_t\equiv m_t^2/M_W^2$, $S_0(x_{t})={\cal O}(1)$, and
\begin{equation}\label{CP-def}
({\cal CP})\left\vert B^{0}_q\right\rangle=
e^{i\phi_{\mbox{{\scriptsize CP}}}(B_q)}
\left\vert\overline{B^{0}_q}\right\rangle.
\end{equation}
Moreover, we obtain from the absorptive parts of the boxes:
\begin{equation}
\frac{\Gamma_{12}^{(q)}}{M_{12}^{(q)}}\approx
-\frac{3\pi}{2S_0(x_{t})}\frac{m_b^2}{M_W^2}={\cal O}(m_b^2/m_t^2)\ll 1.
\end{equation}
Consequently, neglecting 2nd order terms, we arrive at
\begin{equation}
\alpha_q=\left[1+\frac{|\Gamma_{12}^{(q)}|}{2|M_{12}^{(q)}|}\sin\delta
\Theta_{M/\Gamma}^{(q)}\right]e^{-i\left(\Theta_{M_{12}}^{(q)}+n'\pi\right)}.
\end{equation}
The deviation of $|\alpha_q|$ from 1 describes $CP$ violation 
in $B^0_q$--$\overline{B^0_q}$ oscillations, and can be probed through
``wrong-charge'' lepton asymmetries:
\begin{displaymath}
{\cal A}^{(q)}_{\mbox{{\scriptsize SL}}}\equiv
\frac{\Gamma(B^0_q(t)\to l^-\overline{\nu}_l X)-\Gamma(\overline{B^0_q}(t)\to
l^+\nu_l X)}{\Gamma(B^0_q(t)\to l^-\overline{\nu}_l X)+
\Gamma(\overline{B^0_q}(t)\to l^+\nu_l X)}
\end{displaymath}
\begin{equation}\label{wrongcharge}
=\frac{|\alpha_q|^4-1}{|\alpha_q|^4+1}\approx\frac{|\Gamma_{12}^{(q)}|}
{|M_{12}^{(q)}|}\sin\delta\Theta^{(q)}_{M/\Gamma}.
\end{equation}
Note that the time dependences cancel in (\ref{wrongcharge}). Because of 
\begin{equation}
|\Gamma_{12}^{(q)}|/|M_{12}^{(q)}|\propto
m_b^2/m_t^2,\quad
\sin\delta\Theta^{(q)}_{M/\Gamma}\propto m_c^2/m_b^2, 
\end{equation}
the asymmetry ${\cal A}^{(q)}_{\mbox{{\scriptsize SL}}}$ is suppressed by 
a factor $m_c^2/m_t^2={\cal O}(10^{-4})$, and is hence very small in the 
Standard Model. Consequently, it represents an interesting probe to
search for new physics. 

\subsection{Time Evolution}
The time evolution of initially, i.e.\ at $t=0$, pure $|B^0_q\rangle$- and 
$|\overline{B^0_q}\rangle$-meson states is given by
\begin{eqnarray}
\left|B^0_q(t)\right\rangle&=&f_+^{(q)}(t)\left|B^{0}_q\right\rangle
+\alpha_qf_-^{(q)}(t)\left|\overline{B^{0}_q}\right\rangle\\
\left|\overline{B^0_q}(t)\right\rangle&=&\frac{1}{\alpha_q}f_-^{(q)}(t)
\left|B^{0}_q\right\rangle+f_+^{(q)}(t)\left|\overline{B^{0}_q}\right
\rangle,
\end{eqnarray}
where
\begin{equation}
f_{\pm}^{(q)}(t)=\frac{1}{2}\left[e^{-i\lambda_+^{(q)}t}\pm
e^{-i\lambda_-^{(q)}t}\right].
\end{equation}
These time-dependent state vectors allow the calculation of the 
corresponding transition rates. To this end, it is useful to introduce
\begin{equation}
\left|g^{(q)}_{\pm}(t)\right|^2=\frac{1}{4}\left[e^{-\Gamma_{\rm L}^{(q)}t}+
e^{-\Gamma_{\rm H}^{(q)}t}\pm2\,e^{-\Gamma_q t}\cos(\Delta M_qt)\right]
\end{equation}
\begin{equation}
g_-^{(q)}(t)\,g_+^{(q)}(t)^\ast=\frac{1}{4}\left[e^{-\Gamma_{\rm L}^{(q)}t}-
e^{-\Gamma_{\rm H}^{(q)}t}+2\,i\,e^{-\Gamma_q t}\sin(\Delta M_qt)\right],
\end{equation}
and
\begin{equation}\label{xi-def}
\xi_f^{(q)}=e^{-i\Theta_{M_{12}}^{(q)}}
\frac{A(\overline{B_q^0}\to f)}{A(B_q^0\to f)},\quad
\xi_{\overline{f}}^{(q)}=e^{-i\Theta_{M_{12}}^{(q)}}
\frac{A(\overline{B_q^0}\to \overline{f})}{A(B_q^0\to \overline{f})},
\end{equation}
where
\begin{equation}\label{theta-def}
\Theta_{M_{12}}^{(q)}=\pi+2\,\mbox{arg}\left(V_{tq}^\ast V_{tb}\right)-
\phi_{\mbox{{\scriptsize CP}}}(B_q)
\end{equation}
is the $CP$-violating weak $B^0_q$--$\overline{B^0_q}$ mixing phase. 
Whereas $\Theta_{M_{12}}^{(q)}$ depends on the chosen CKM and $CP$ 
phase conventions, $\xi_f^{(q)}$ and $\xi_{\overline{f}}^{(q)}$ are 
{\it convention-independent observables}.

The $g_{\pm}^{(q)}(t)$ are related to the $f_{\pm}^{(q)}(t)$. However, 
whereas the latter functions depend on $n'$, the $g_{\pm}^{(q)}(t)$ do 
not depend on this parameter. The $n'$-dependence is cancelled by 
introducing the {\it positive} mass difference 
\begin{equation}
\Delta M_q\equiv M_{\rm H}^{(q)}-M_{\rm L}^{(q)}=2\left|M_{12}^{(q)}\right|>0
\end{equation}
of the mass eigenstates $|B_q^{\rm H}\rangle$ (``heavy'') and
$|B_q^{\rm L}\rangle$ (``light''). The quantities $\Gamma_{\rm H}^{(q)}$ and 
$\Gamma_{\rm L}^{(q)}$ denote the corresponding decay widths. 
Their difference can be expressed as
\begin{equation}
\Delta\Gamma_q\equiv\Gamma_{\rm H}^{(q)}-\Gamma_{\rm L}^{(q)}=
\frac{4\mbox{\,Re}\left[M_{12}^{(q)}\Gamma_{12}^{(q)\ast}\right]}{\Delta M_q},
\end{equation}
whereas their average is given by
\begin{equation}
\Gamma_q\equiv\frac{\Gamma^{(q)}_{\rm H}+\Gamma^{(q)}_{\rm L}}{2}=
\Gamma^{(q)}_0.
\end{equation}
There is the following interesting relation:
\begin{equation}
\frac{\Delta\Gamma_q}{\Gamma_q}\approx-\frac{3\pi}{2S_0(x_t)}\frac{m_b^2}
{M_W^2}\,x_q={\cal O}(10^{-2})\times x_q,
\end{equation}
where
\begin{equation}\label{mix-par}
x_q\equiv\frac{\Delta M_q}{\Gamma_q}=\left\{\begin{array}{cc}
0.723\pm0.032&(q=d)\\
{\cal O}(20)& (q=s)
\end{array}\right.
\end{equation}
denotes the $B^0_q$--$\overline{B^0_q}$ ``mixing parameter''. Consequently,
there may be a sizeable width difference in the $B_s$ system, whereas 
$\Delta\Gamma_d$ is expected to be negligibly small. We shall come back to 
$\Delta\Gamma_s$ in Section~\ref{sec:Bs}.

Combining the formulae listed above, we arrive at the following
transition rates for decays of initially, i.e.\ at $t=0$, present $B^0_q$-
and $\overline{B^0_q}$-mesons:
\begin{eqnarray}
\lefteqn{\Gamma(\stackrel{{\mbox{\tiny (---)}}}{B^0_q}(t)\to f)}\nonumber\\
&&=\left[\left|g_\mp^{(q)}(t)\right|^2+\left|\xi_f^{(q)}
\right|^2\left|g_\pm^{(q)}(t)\right|^2-
2\mbox{\,Re}\left\{\xi_f^{(q)}
g_\pm^{(q)}(t)g_\mp^{(q)}(t)^\ast\right\}
\right]\tilde\Gamma_f,\label{rates}
\end{eqnarray}
where the time-independent rate $\tilde\Gamma_f$ corresponds to the 
``unevolved'' decay amplitude $A(B^0_q\to f)$, which can be calculated by 
performing the usual phase-space integrations. The rates into the
$CP$-conjugate final state $\overline{f}$ can be
obtained straightforwardly from (\ref{rates}) through the substitutions
\begin{equation}
\tilde\Gamma_f  \,\,\,\to\,\,\, 
\tilde\Gamma_{\overline{f}},
\quad\,\,\xi_f^{(q)} \,\,\,\to\,\,\, 
\xi_{\overline{f}}^{(q)}.
\end{equation}

\subsection{CP-violating Asymmetries}\label{subsec:CPasym}
A particularly simple and interesting situation arises if we restrict 
ourselves to decays of neutral $B_q$-mesons into $CP$ self-conjugate 
final states $|f\rangle$, \mbox{satisfying} the relation 
\begin{equation}
({\cal CP})|f\rangle=\pm\,|f\rangle. 
\end{equation}
Consequently, we have $\xi_f^{(q)}=\xi_{\overline{f}}^{(q)}$ in this
case (see (\ref{xi-def})). Using (\ref{rates}), the corresponding 
time-dependent $CP$ asymmetry can be expressed as
\begin{eqnarray}
\lefteqn{a_{\rm CP}(t)\equiv\frac{\Gamma(B^0_q(t)\to f)-
\Gamma(\overline{B^0_q}(t)\to f)}{\Gamma(B^0_q(t)\to f)+
\Gamma(\overline{B^0_q}(t)\to f)}}\label{ee6}\\
&&=\left[\frac{{\cal A}_{\rm CP}^{\rm dir}(B_q\to f)\,\cos(\Delta M_q t)+
{\cal A}_{\rm CP}^{\rm mix}(B_q\to f)\,\sin(\Delta 
M_q t)}{\cosh(\Delta\Gamma_qt/2)-{\cal A}_{\rm 
\Delta\Gamma}(B_q\to f)\,\sinh(\Delta\Gamma_qt/2)}\right],\nonumber
\end{eqnarray}
where we have separated the ``direct'' from the ``mixing-induced'' 
$CP$-violating contributions, which are described by
\begin{equation}\label{ee7}
{\cal A}^{\mbox{{\scriptsize dir}}}_{\mbox{{\scriptsize CP}}}(B_q\to f)\equiv
\frac{1-\bigl|\xi_f^{(q)}\bigr|^2}{1+\bigl|\xi_f^{(q)}\bigr|^2}\quad
\mbox{and}\quad
{\cal A}^{\mbox{{\scriptsize mix}}}_{\mbox{{\scriptsize
CP}}}(B_q\to f)\equiv\frac{2\,\mbox{Im}\,\xi^{(q)}_f}{1+
\bigl|\xi^{(q)}_f\bigr|^2}\,,
\end{equation}
respectively. The terminology ``direct $CP$ violation'' refers 
to $CP$-violating effects, which arise directly in the corresponding 
decay amplitudes and are due to interference between different CKM 
amplitudes. On the other hand, ``mixing-induced $CP$ violation'' originates
from interference effects between $B_q^0$--$\overline{B_q^0}$ mixing 
and decay processes. The width difference $\Delta\Gamma_q$, which may 
be sizeable in the $B_s$ system, provides another observable 
\begin{equation}\label{ADGam}
{\cal A}_{\rm \Delta\Gamma}(B_q\to f)\equiv
\frac{2\,\mbox{Re}\,\xi^{(q)}_f}{1+\bigl|\xi^{(q)}_f
\bigr|^2},
\end{equation}
which is, however, not independent from ${\cal A}^{\mbox{{\scriptsize 
dir}}}_{\mbox{{\scriptsize CP}}}(B_q\to f)$ and 
${\cal A}^{\mbox{{\scriptsize mix}}}_{\mbox{{\scriptsize CP}}}(B_q\to f)$:
\begin{equation}\label{Obs-rel}
\Bigl[{\cal A}_{\rm CP}^{\rm dir}(B_q\to f)\Bigr]^2+
\Bigl[{\cal A}_{\rm CP}^{\rm mix}(B_q\to f)\Bigr]^2+
\Bigl[{\cal A}_{\Delta\Gamma}(B_q\to f)\Bigr]^2=1.
\end{equation}

In order to calculate the observable $\xi_f^{(q)}$, containing 
essentially all the information needed to evaluate the $CP$ asymmetry
(\ref{ee6}), we employ the low-energy effective Hamiltonian (\ref{e4}):
\begin{eqnarray}
\lefteqn{A\left(\overline{B^0_q}\to f\right)=\Bigl\langle f\Bigl\vert
{\cal H}_{\mbox{{\scriptsize eff}}}(\Delta B=-1)\Bigr\vert\overline{B^0_q}
\Bigr\rangle=}\\
&&\Biggl\langle f\left|
\frac{G_{\mbox{{\scriptsize F}}}}{\sqrt{2}}\left[
\sum\limits_{j=u,c}V_{jr}^\ast V_{jb}\left\{\sum\limits_{k=1}^2
C_{k}(\mu)\,Q_{k}^{jr}(\mu)
+\sum\limits_{k=3}^{10}C_{k}(\mu)\,Q_{k}^r(\mu)\right\}\right]\right|
\overline{B^0_q}\Biggr\rangle,~~~\mbox{}\nonumber
\end{eqnarray}
where $r\in\{d,s\}$ distinguishes between $b\to d$ and $b\to s$ transitions.
On the other hand, we also have 
\begin{eqnarray}
\lefteqn{A\left(B^0_q\to f\right)=\left\langle f\left|
{\cal H}_{\mbox{{\scriptsize 
eff}}}(\Delta B=-1)^\dagger\right|B^0_q\right\rangle=}\\
&&\hspace*{-0.5truecm}\Biggl\langle f
\left|\frac{G_{\mbox{{\scriptsize F}}}}{\sqrt{2}}
\left[\sum\limits_{j=u,c}V_{jr}V_{jb}^\ast \left\{\sum\limits_{k=1}^2
C_{k}(\mu)\,Q_{k}^{jr\dagger}(\mu)+\sum\limits_{k=3}^{10}
C_{k}(\mu)\,Q_k^{r\dagger}(\mu)\right\}\right]\right|B^0_q
\Biggr\rangle.~~~\mbox{}\nonumber
\end{eqnarray}
Performing appropriate $CP$ transformations in this expression, i.e.\ 
inserting the operator $({\cal CP})^\dagger({\cal CP})=\hat 1$ both 
after $\langle f|$ and in front of $|B^0_q\rangle$, yields
\begin{eqnarray}
\lefteqn{A\left(B^0_q\to f\right)=\pm e^{i\phi_{\mbox{{\scriptsize CP}}}
(B_q)}\times}\\
&&\Biggl\langle f\left|
\frac{G_{\mbox{{\scriptsize F}}}}{\sqrt{2}}\left[\sum\limits_{j=u,c}
V_{jr}V_{jb}^\ast\left\{\sum\limits_{k=1}^2
C_{k}(\mu)\,Q_{k}^{jr}(\mu)+\sum\limits_{k=3}^{10}
C_{k}(\mu)\,Q_{k}^r(\mu)\right\}\right]\right|\overline{B^0_q}
\Biggr\rangle,\nonumber
\end{eqnarray}
where we have applied the relation 
$({\cal CP})Q_k^{jr\dagger}({\cal CP})^\dagger=Q_k^{jr}$, and have 
furthermore taken into account (\ref{CP-def}). Using now (\ref{xi-def})
and (\ref{theta-def}), we finally arrive at
\begin{equation}\label{xi-expr}
\xi_f^{(q)}=\mp\,e^{-i\phi_q}\,
\left[\frac{\sum\limits_{j=u,c}V_{jr}^\ast V_{jb}\bigl\langle 
f\bigl|{\cal Q}^{jr}\bigr|\overline{B^0_q}\bigr\rangle}{\sum\limits_{j=u,c}
V_{jr}V_{jb}^\ast\bigl\langle f\bigl|{\cal Q}^{jr}
\bigr|\overline{B^0_q}\bigr\rangle}\right],
\end{equation}
where 
\begin{equation}
{\cal Q}^{jr}\equiv\sum\limits_{k=1}^2C_k(\mu)\,Q_k^{jr}+
\sum\limits_{k=3}^{10}C_k(\mu)\,Q_k^{r},
\end{equation}
and where
\begin{equation}
\phi_q\equiv 2\,\mbox{arg} (V_{tq}^\ast V_{tb})=\left\{\begin{array}{cr}
+2\beta&\mbox{($q=d$)}\\
-2\delta\gamma&\mbox{($q=s$)}\end{array}\right.
\end{equation}
is related to the weak $B_q^0$--$\overline{B_q^0}$ mixing phase. Note that
the phase-convention-dependent quantity 
$\phi_{\mbox{{\scriptsize CP}}}(B_q)$ cancels in this expression. 

In general, the observable $\xi_f^{(q)}$ suffers from large hadronic 
uncertainties, which are introduced by the hadronic matrix elements in 
Eq.~(\ref{xi-expr}). However, if the decay $B_q\to f$ is dominated by a 
single CKM amplitude, i.e.
\begin{equation}
A(B^0_q\to f)=e^{-i\phi_f/2}
\left[e^{i\delta_f}|M_f|\right],
\end{equation}
the strong matrix element $e^{i\delta_f}|M_f|$ cancels, and $\xi_f^{(q)}$ 
takes the simple form
\begin{equation}\label{ee10}
\xi_f^{(q)}=\mp\exp\left[-i\left(\phi_q-\phi_f\right)\right].
\end{equation}
If the $V_{jr}^\ast V_{jb}$ amplitude plays the dominant role in 
$\overline{B^0_q}\to f$, we have
\begin{equation}
\phi_f=2\,\mbox{arg}(V_{jr}^\ast V_{jb})=\left\{\begin{array}{cc}
-2\gamma&\mbox{($j=u$)}\\
0&\,\mbox{($j=c$).}
\end{array}\right.
\end{equation}

\section{Important B-Factory Benchmark Modes}\label{sec:benchmark}
The formalism discussed in Subsection~\ref{subsec:CPasym}
has several interesting applications. The most important one is the 
extraction of the CKM angle $\beta$ from $CP$-violating effects in the 
``gold-plated'' mode $B_d\to J/\psi\,K_{\rm S}$ \cite{bisa}.

\subsection{Extracting $\beta$ from 
$B^0\to J/\psi K_{\rm S}$}\label{subsec:bpsiks}
\begin{figure}
\begin{center}
\leavevmode
\epsfysize=4.4truecm 
\epsffile{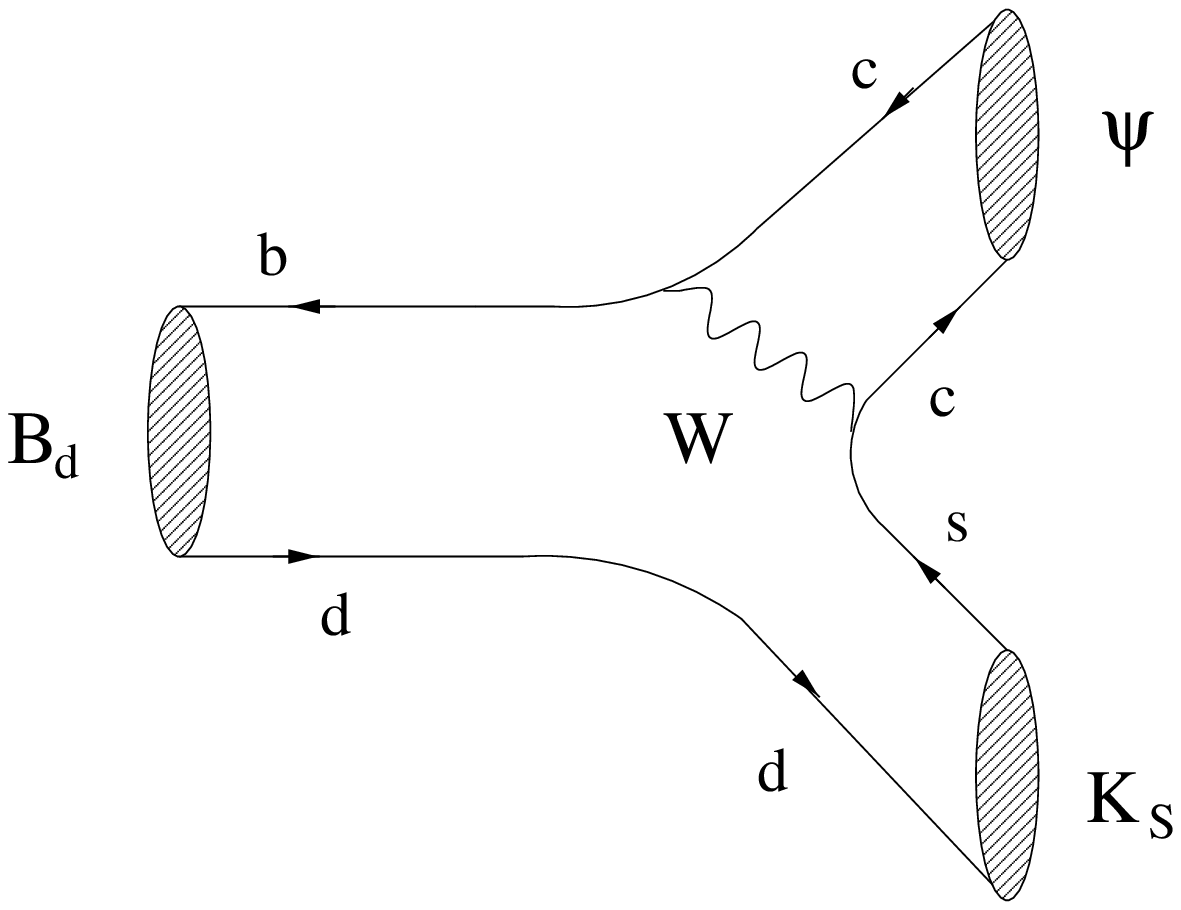} \hspace*{1truecm}
\epsfysize=4.4truecm 
\epsffile{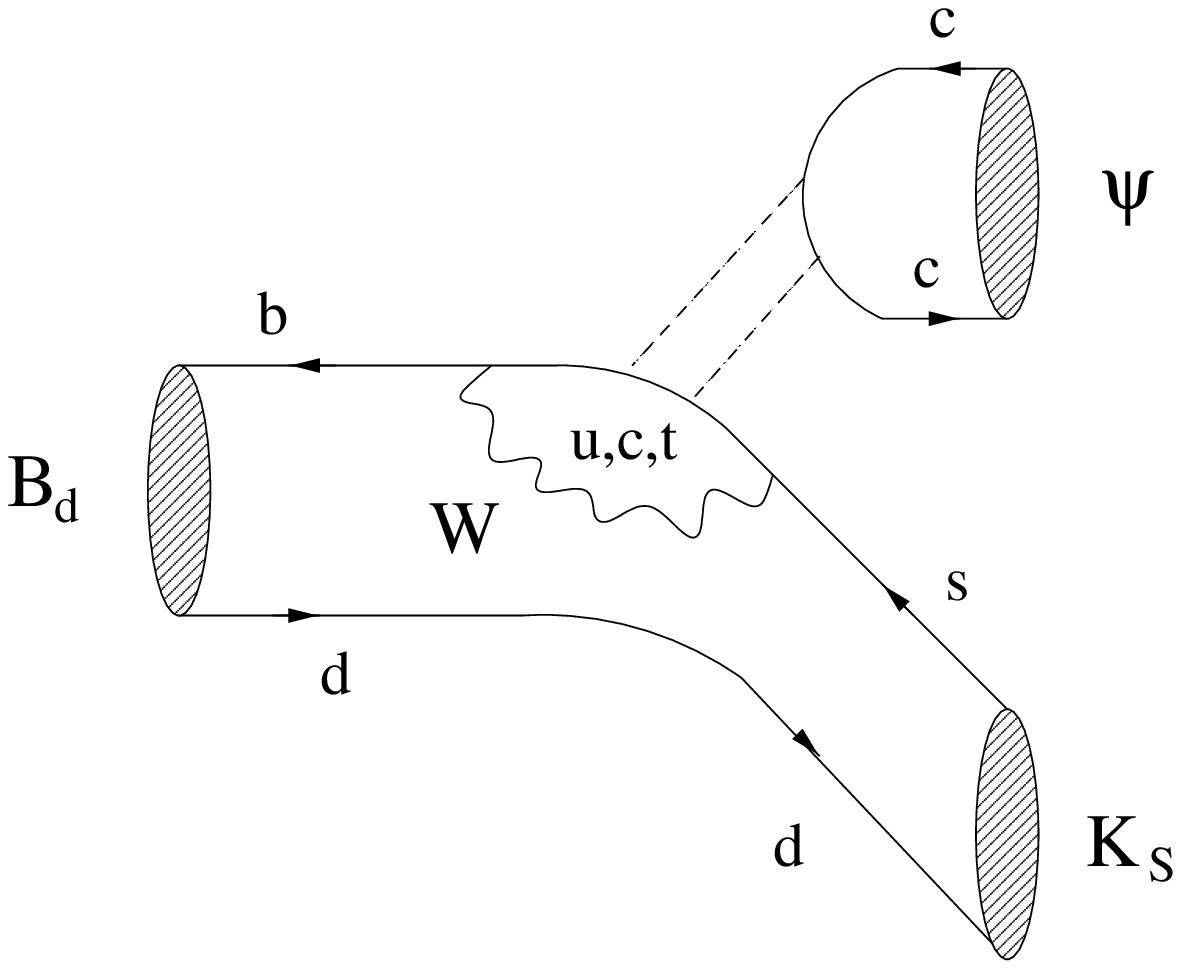}
\end{center}
\vspace*{-0.3truecm}
\caption{Feynman diagrams contributing to $B_d^0\to J/\psi\,K_{\rm S}$.
The dashed lines in the penguin topology represent a colour-singlet 
exchange.}\label{fig:BdPsiKS}
\end{figure}

The decay $B_d^0\to J/\psi\,K_{\rm S}$ is a transition into a $CP$ eigenstate 
with eigenvalue $-1$, and originates from $\overline{b}\to
\overline{c}\,c\,\overline{s}$ quark-level decays. As can be seen in 
Fig.~\ref{fig:BdPsiKS}, we have to 
deal both with tree-diagram-like and with penguin topologies. The 
corresponding amplitude can be written as \cite{RF-BdsPsiK}
\begin{equation}\label{Bd-ampl1}
A(B_d^0\to J/\psi\, K_{\rm S})=\lambda_c^{(s)}\left(A_{\rm cc}^{c'}+
A_{\rm pen}^{c'}\right)+\lambda_u^{(s)}A_{\rm pen}^{u'}
+\lambda_t^{(s)}A_{\rm pen}^{t'}\,,
\end{equation}
where $A_{\rm cc}^{c'}$ denotes the current--current contributions,
i.e.\ the ``tree'' processes in Fig.\ \ref{fig:BdPsiKS}, and the strong
amplitudes $A_{\rm pen}^{q'}$ describe the contributions from penguin 
topologies with internal $q$ quarks ($q\in\{u,c,t\})$. These penguin 
amplitudes take into account both QCD and electroweak penguin contributions. 
The primes in (\ref{Bd-ampl1}) remind us that we are dealing with a 
$\overline{b}\to\overline{s}$ transition, and the
\begin{equation}\label{lamqs-def}
\lambda_q^{(s)}\equiv V_{qs}V_{qb}^\ast
\end{equation}
are CKM factors. If we make use both of the unitarity of the CKM matrix,
implying $\lambda_u^{(s)}+\lambda_c^{(s)}+\lambda_t^{(s)}=0$,
and of the Wolfenstein parametrization \cite{wolf}, generalized to
include non-leading terms in $\lambda$ \cite{blo}, we obtain 
\begin{equation}\label{Bd-ampl2}
A(B_d^0\to J/\psi\, K_{\rm S})=\left(1-\frac{\lambda^2}{2}\right){\cal A'}
\left[1+\left(\frac{\lambda^2}{1-\lambda^2}\right)a'e^{i\theta'}e^{i\gamma}
\right],
\end{equation}
where
\begin{equation}\label{Aap-def}
{\cal A'}\equiv\lambda^2A\left(A_{\rm cc}^{c'}+A_{\rm pen}^{ct'}\right),
\end{equation}
with $A_{\rm pen}^{ct'}\equiv A_{\rm pen}^{c'}-A_{\rm pen}^{t'}$, and
\begin{equation}\label{ap-def}
a'e^{i\theta'}\equiv R_b\left(\frac{A_{\rm pen}^{ut'}}{A_{\rm cc}^{c'}+
A_{\rm pen}^{ct'}}\right).
\end{equation}
The quantity $A_{\rm pen}^{ut'}$ is defined in analogy to $A_{\rm pen}^{ct'}$,
$A$ is given by (see (\ref{set-rel}))
\begin{equation}\label{CKM-exp}
A=\left|V_{cb}\right|/\lambda^2=0.81\pm0.06\,,
\end{equation}
and the definition of $R_b=0.41\pm0.07$ can be found in (\ref{Rb-def}).

It is very difficult to calculate the ``penguin'' parameter $a'e^{i\theta'}$,
which introduces the $CP$-violating phase factor $e^{i\gamma}$ into
the $B_d^0\to J/\psi\, K_{\rm S}$ decay amplitude and represents -- sloppily 
speaking -- the ratio of the penguin to tree contributions. However, this 
parameter -- and therefore also $e^{i\gamma}$ -- enters (\ref{Bd-ampl2}) 
in a doubly Cabibbo-suppressed way. Consequently, we have to a very good
approximation $\phi_{\psi K_{\rm S}}=0$, and obtain with the help 
of (\ref{ee7}) and~(\ref{ee10}):
\begin{equation}\label{e12}
{\cal A}^{\mbox{{\scriptsize mix}}}_{\mbox{{\scriptsize
CP}}}(B_d\to J/\psi\, K_{\mbox{{\scriptsize S}}})=+\sin[-(\phi_d-0)]=
-\sin(2\beta)\,.
\end{equation}
Since (\ref{ee10}) applies with excellent accuracy to $B_d\to J/\psi\, 
K_{\mbox{{\scriptsize S}}}$, it is referred to as the 
``gold-plated'' mode to determine $\beta$ \cite{bisa}. In addition to 
(\ref{e12}), another important implication of the Standard Model is the 
following relation: 
\begin{equation}\label{e13}
{\cal A}^{\mbox{{\scriptsize dir}}}_{\mbox{{\scriptsize
CP}}}(B_d\to J/\psi\, K_{\mbox{{\scriptsize S}}})\approx0\approx
{\cal A}_{\mbox{{\scriptsize CP}}}(B^+\to J/\psi\, K^+).
\end{equation}
An observation of these $CP$ asymmetries at the level of 10\% would 
be a strong indication for new physics. There is already an interesting
constraint from CLEO \cite{cleo-dir},
${\cal A}_{\mbox{{\scriptsize CP}}}(B^+\to J/\psi\, K^+)=(-1.8\pm4.3\pm0.4)\%$,
and BaBar reported 
${\cal A}^{\mbox{{\scriptsize dir}}}_{\mbox{{\scriptsize
CP}}}(B_d\to J/\psi\, K_{\mbox{{\scriptsize S}}})=\left(26\pm19\right)\%$
\cite{babar}. 

Concerning the measurement of $\sin(2\beta)$ through (\ref{e12}),
there were already important first steps by the OPAL, CDF and ALEPH
collaborations:
\begin{equation}
\sin(2\beta)=\left\{\begin{array}{ll}
3.2^{+1.8}_{-2.0}\pm0.5&\mbox{(OPAL \cite{opal})}\\
0.79^{+0.41}_{-0.44}&\mbox{(CDF \cite{cdf})}\\
0.84^{+0.82}_{-1.04}\pm0.16&\mbox{(ALEPH \cite{aleph}).}
\end{array}\right.
\end{equation}
In the summer of 2000, also the first results from the 
asymmetric $e^+$--$e^-$ $B$-factories were reported:
\begin{equation}\label{B-factory}
\sin(2\beta)=\left\{\begin{array}{ll}
0.12\pm0.37\pm0.09&\,\mbox{(BaBar \cite{babar})}\\
0.45^{+0.43+0.07}_{-0.44-0.09}&\,\mbox{(Belle \cite{belle}).}
\end{array}\right.
\end{equation}
On the other hand, the CKM fits discussed in Subsection~\ref{subsec:CKM-fits} 
yield the following range for the Standard-Model expectation \cite{AL}:
\begin{equation}\label{SM-sin}
0.53\leq\sin(2\beta)\leq0.93.
\end{equation}
Although the experimental uncertainties are still very large, the
small central value reported by the BaBar collaboration 
\cite{babar} led already to some excitement in the $B$-physics community 
\cite{low-sin}, as it would be in conflict with the Standard-Model 
range (\ref{SM-sin}). This possible 
discrepancy might indicate new-physics contributions to 
$B^0_d$--$\overline{B^0_d}$ and/or $K^0$--$\overline{K^0}$ mixing. 

After a couple of years collecting data at the $B$-factories, an experimental 
uncertainty of $\left.\Delta\sin(2\beta)\right|_{\rm exp}=0.05$ seems to be 
achievable, whereas the experimental uncertainty in the LHC era is expected 
to be one order of magnitude higher \cite{LHC-Report}. In view of this 
tremendous experimental accuracy, it is an important issue to investigate 
the theoretical accuracy of (\ref{e12}) and (\ref{e13}), which is a very 
challenging theoretical task. An interesting channel in this respect is 
$B_s\to J/\psi\,K_{\rm S}$ \cite{RF-BdsPsiK}, allowing us to control the -- 
presumably very small -- penguin uncertainties in the determination of 
$\beta$ from the $CP$-violating effects in $B_d\to J/\psi\,K_{\rm S}$, and 
to extract the CKM angle $\gamma$. 

\begin{figure}
\begin{center}
\leavevmode
\epsfysize=4.0truecm 
\epsffile{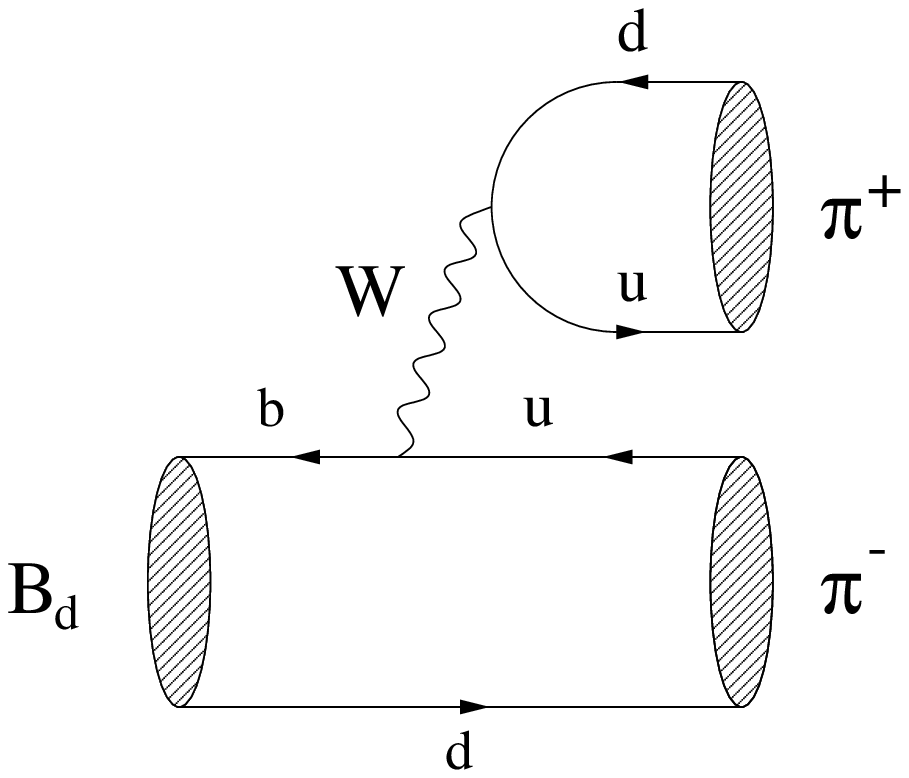} \hspace*{1truecm}
\epsfysize=4.5truecm 
\epsffile{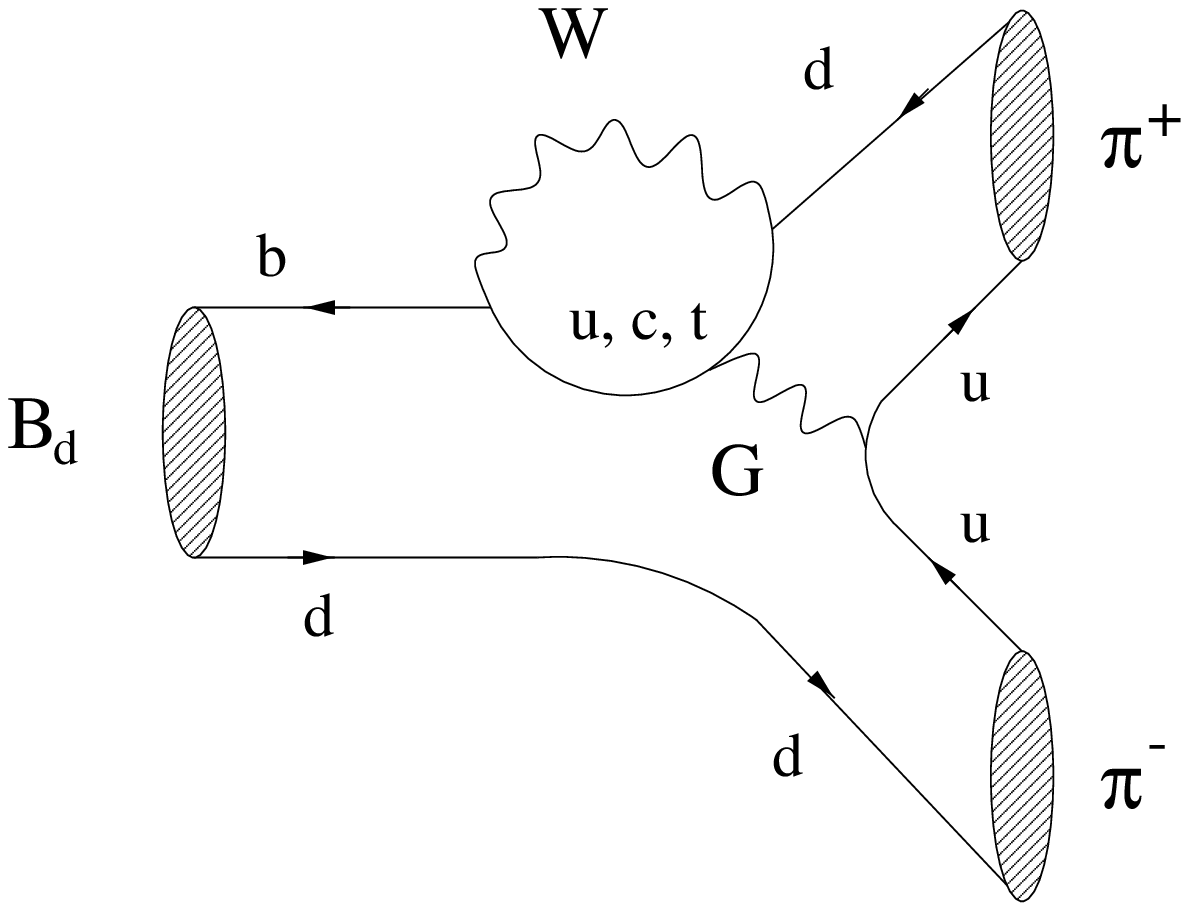}
\end{center}
\vspace*{-0.3truecm}
\caption{Feynman diagrams contributing to 
$B_d^0\to\pi^+\pi^-$.}\label{fig:bpipi}
\end{figure}

\subsection{Probing $\alpha$ through $B^0\to\pi^+\pi^-$}\label{BPIPI-alpha}
Another benchmark mode for the $B$-factories is 
$B_d^0\to\pi^+\pi^-$. It is a decay into a $CP$ eigenstate with eigenvalue 
$+1$, and originates from $\overline{b}\to\overline{u}\,u\,\overline{d}$ 
quark-level transitions (see Fig.~\ref{fig:bpipi}). In analogy to 
(\ref{Bd-ampl1}), the corresponding decay amplitude can be expressed in the 
following way \cite{RF-BsKK}:
\begin{eqnarray}
\lefteqn{A(B_d^0\to\pi^+\pi^-)=
\lambda_u^{(d)}\left(A_{\rm cc}^{u}+
A_{\rm pen}^{u}
\right)+\lambda_c^{(d)}A_{\rm pen}^{c}+\lambda_t^{(d)}A_{\rm pen}^{t}}
\qquad\qquad\qquad\mbox{}\nonumber\\
&&=\left(1-\frac{\lambda^2}{2}\right){\cal C}
\left[e^{i\gamma}-d\,e^{i\theta}\right],\label{Bpipi-ampl}
\qquad\qquad\mbox{}
\end{eqnarray}
where
\begin{equation}\label{C-DEF}
{\cal C}\equiv\lambda^3A\,R_b\left(A_{\rm cc}^{u}+A_{\rm pen}^{ut}\right)
\quad\mbox{with}\quad A_{\rm pen}^{ut}\equiv A_{\rm pen}^{u}-A_{\rm pen}^{t},
\end{equation}
and
\begin{equation}\label{D-DEF}
d\,e^{i\theta}\equiv\frac{1}{(1-\lambda^2/2)R_b}
\left(\frac{A_{\rm pen}^{ct}}{A_{\rm cc}^{u}+A_{\rm pen}^{ut}}\right).
\end{equation}
In contrast to the $B_d^0\to J/\psi\, K_{\mbox{{\scriptsize S}}}$ amplitude
(\ref{Bd-ampl2}), the ``penguin'' parameter $d\,e^{i\theta}$ does 
{\it not} enter (\ref{Bpipi-ampl}) in a doubly Cabibbo-suppressed way.
If we assume, for a moment, that $d=0$, the formalism discussed in 
Subsection~\ref{subsec:CPasym} yields
\begin{equation}\label{bpipi-ideal}
{\cal A}^{\mbox{{\scriptsize mix}}}_{\mbox{{\scriptsize
CP}}}(B_d\to\pi^+\pi^-)=-\sin[-(2\beta+2\gamma)]=-\sin(2\alpha),
\end{equation}
which would allow a determination of $\alpha$. However, theoretical 
estimates typically give $d={\cal O}(0.2)$, and also the present CLEO
data on $B\to\pi K$ modes indicate that penguins play in fact an
important role \cite{RF-bpipi}. Consequently, the approximation $d=0$, 
i.e.\ the neglect of penguins in $B_d\to\pi^+\pi^-$, is not justified. 
The penguin uncertainties affecting (\ref{bpipi-ideal}) 
were analysed by many authors during the last couple of years 
\cite{BBNS,alpha-uncert}. 

\begin{figure}
\centerline{
\epsfxsize=7.8truecm
\epsffile{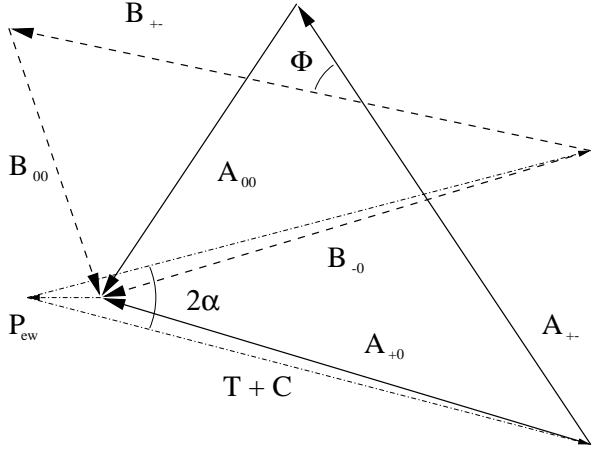}}
\caption{The $B\to\pi\pi$ isospin triangles in the complex 
plane. Here the amplitudes $A$ correspond to (\ref{su21});
the amplitudes $B$ correspond to the ones in (\ref{su22}), rotated by 
$e^{-2i\beta}$.}\label{fig:bpipi-isospin}
\end{figure}

There are strategies to control the penguin uncertainties with the help 
of additional experimental data. The best known approach was proposed by 
Gronau and London \cite{GL}, and makes use of isospin relations between the
$B\to\pi\pi$ decay amplitudes. Since $B^\pm\to\pi^\pm\pi^0$ is a $\Delta I=3/2$
transition, the QCD penguin operators (\ref{qcd-penguins}), which mediate 
$\Delta I=1/2$ transitions, do not contribute. Consequently, 
if we neglect EW penguins for a moment, we obtain
\begin{equation}
A(B^\pm\to\pi^\pm\pi^0)=e^{\pm i\gamma}e^{i\delta{T+C}}|T+C|,
\end{equation}
which yields
\begin{equation}\label{rel1}
\frac{A(B^+\to\pi^+\pi^0)}{A(B^-\to\pi^-\pi^0)}=e^{2i\gamma}.
\end{equation}
Moreover, the isospin symmetry implies the following amplitude relations:
\begin{eqnarray}
\sqrt{2}\,A(B^+\to\pi^+\pi^0)&=&A(B^0_d\to\pi^+\pi^-)+
\sqrt{2}\,A(B^0_d\to\pi^0\pi^0)\label{su21}\\
\sqrt{2}\,A(B^-\to\pi^-\pi^0)&=&A(\overline{B^0_d}\to\pi^+\pi^-)+
\sqrt{2}\,A(\overline{B^0_d}\to\pi^0\pi^0),\label{su22}
\end{eqnarray}
which can be represented as two triangles in the complex plane. These
triangles can be fixed through the measured six $B\to\pi\pi$ branching
ratios. In order to determine their relative orientation, we rotate
the $CP$-conjugate triangle by $e^{-2i\beta}$. The corresponding situation
is illustrated in Fig.~\ref{fig:bpipi-isospin}, where the angle $\Phi$
can be fixed through mixing-induced $CP$ violation \cite{PAPI}:
\begin{equation}
{\cal A}^{\mbox{{\scriptsize mix}}}_{\mbox{{\scriptsize
CP}}}(B_d\to\pi^+\pi^-)=-\,\frac{2|A_{+-}||B_{+-}|}{|A_{+-}|^2+
|B_{+-}|^2}\sin\Phi\,.
\end{equation}
Using (\ref{rel1}), and taking into account that the $CP$-conjugate triangle
was rotated by $e^{-2i\beta}$, we conclude that the angle between the
$B^+\to\pi^+\pi^0$ and $B^-\to\pi^-\pi^0$ amplitudes is given by $2\alpha$.
The EW penguin amplitude $P_{\rm ew}$, which was neglected so far, affects 
this determination of $\alpha$, as can be seen in Fig.~\ref{fig:bpipi-isospin}.
Although EW penguins play a minor role in this construction, they
can be taken into account with the help of the $SU(2)$ isospin symmetry,
which implies \cite{BF-BpiK1,GPY}
\begin{equation}
\left[\frac{P_{\rm ew}}{T+C}\right]=
-\,1.3\times 10^{-2}\times\frac{|V_{td}|}{|V_{ub}|}\,e^{i\alpha}.
\end{equation}

Unfortunately, the $B\to\pi\pi$ triangle approach is very challenging from
an experimental point of view because of 
BR$\left.(B_d\to\pi^0\pi^0)\right|_{\rm TH}
\mathrel{\hbox{\rlap{\hbox{\lower4pt\hbox{$\sim$}}}\hbox{$<$}}}
{\cal O}(10^{-6})$. 
Therefore, alternative strategies are needed. An important one is provided by 
$B\to\rho\pi$ modes \cite{Brhopi}. Here the final states consist of $I=0,1,2$ 
configurations, and the isospin symmetry implies two pentagonal relations,
which correspond to (\ref{su21}) and (\ref{su22}), and also allow a 
determination of $\alpha$. This approach is quite complicated. 
However, it can be simplified by considering a maximum-likelihood fit to the 
parameters of the full Dalitz plot distribution, where it is assumed 
that the $B\to3\pi$ events are fully dominated by $B\to\rho\pi$ \cite{qs}. 
Further simplifications were proposed in Ref.~\cite{qusi}. An issue raised 
recently in this context is the impact of ``polar diagrams'', yielding
$B\to\{\pi(B^\ast,\rho)\}\to\pi\pi\pi$ transitions. These processes may affect 
$B^\mp\to\pi^\mp\pi^\mp\pi^\pm$ and $B_d\to\rho^0\pi^0$, and represent 
an irreducible background in the Dalitz plot \cite{dea}. 

Another possibility to eliminate the penguin uncertainties in the
extraction of $\alpha$ from $B_d\to\pi^+\pi^-$ is to combine this
channel with $B_d\to K^0\overline{K^0}$ through the $SU(3)$ flavour
symmetry \cite{BF-alpha}. A simple strategy to extract $\alpha$, making
also use of $SU(3)$, was proposed in \cite{FM-alpha}; refinements of this 
approach and further interesting methods were suggested in \cite{charles}.

Let us finally note that a particularly interesting strategy is provided 
by the decay $B_s\to K^+K^-$, which is related to $B_d\to\pi^+\pi^-$ by
interchanging all down and strange quarks, i.e.\ through the $U$-spin
flavour symmetry of strong interactions. A combined analysis of these two 
channels allows a simultaneous determination of $\beta$ and $\gamma$ 
\cite{RF-BsKK}, which has certain theoretical advantages, appears to be 
promising for CDF-II \cite{wuerth}, and is ideally suited for LHCb 
\cite{LHC-Report}. This approach is discussed in detail in 
Section~\ref{sec:Uspin}.

\begin{figure}
\begin{center}
\leavevmode
\epsfysize=4.0truecm 
\epsffile{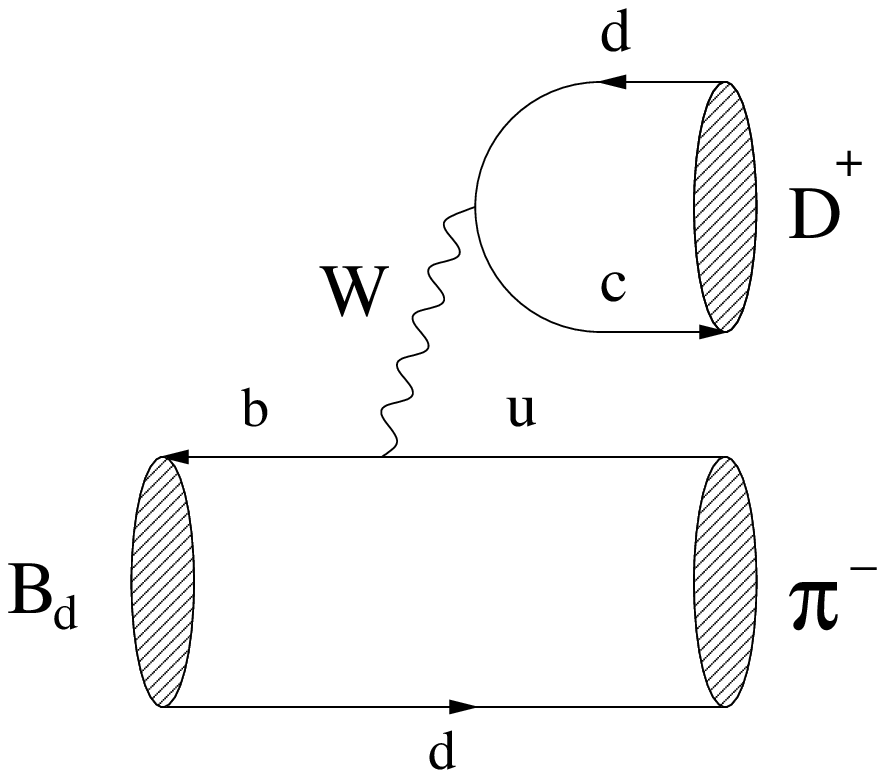} \hspace*{1truecm}
\epsfysize=4.0truecm 
\epsffile{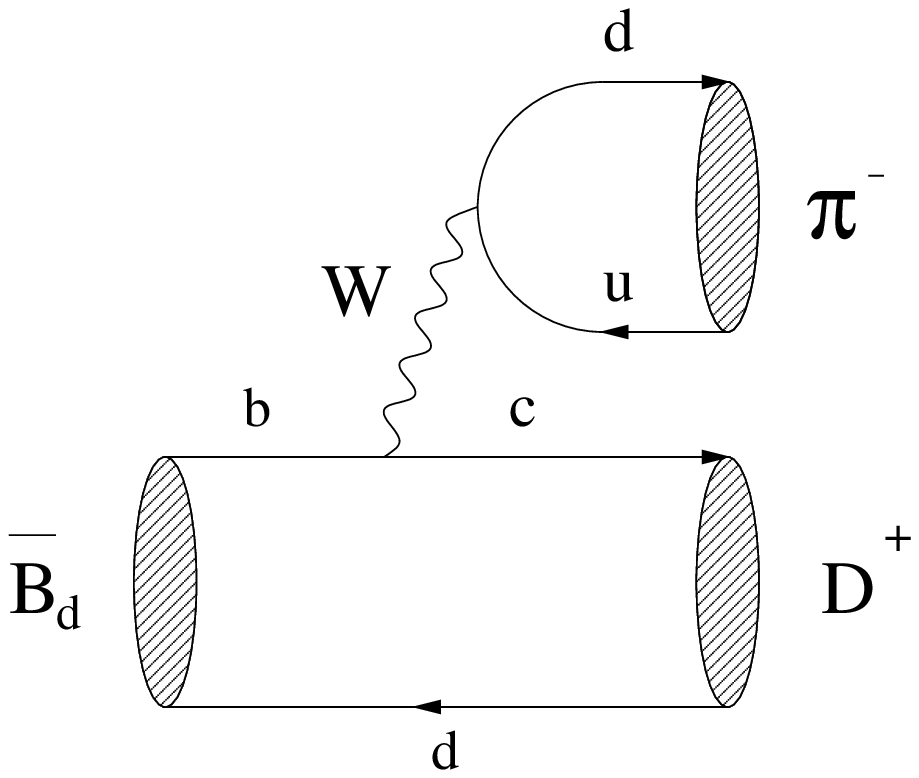}
\end{center}
\vspace*{-0.4truecm}
\caption[]{Feynman diagrams contributing to $B^0_d, \overline{B^0_d}\to 
D^{(\ast)+}\pi^-$.}\label{fig:BDpi}
\end{figure}

\subsection{Extracting $2\beta+\gamma$ from $B^0\to D^{(\ast)\pm}\pi^{\mp}$
Decays}\label{subsec:BDpi}
So far, we have put a strong emphasis on neutral $B$ decays into final 
$CP$ eigenstates. However, in order to extract CKM phases, there are also 
interesting decays of $B_{d,s}$ mesons into final states that are {\it not} 
eigenstates of the $CP$ operator. An important example is given by
$B_d\to D^{(\ast)\pm}\pi^{\mp}$ decays, which receive only contributions 
from tree-diagram-like topologies. As can be seen in Fig.~\ref{fig:BDpi}, 
$B^0_d$- and $\overline{B^0_d}$-mesons may both decay into 
$D^{(\ast)+}\pi^-$, thereby leading to interference 
effects between $B^0_d$--$\overline{B^0_d}$ mixing and decay processes. 
Consequently, the time-dependent rates for decays of initially, i.e.\ at 
time $t=0$, present $B^0_d$- or $\overline{B^0_d}$-mesons into the 
final state $f\equiv D^{(\ast)+}\pi^-$ (see (\ref{rates})) allow us to 
determine the observable \cite{RF-rev}
\begin{equation}
\xi_f^{(d)}=e^{-i\Theta_{M_{12}}^{(d)}}
\frac{A(\overline{B^0_d}\to f)}{A(B^0_d\to f)}
=-\,e^{-i(\phi_d+\gamma)}\left(\frac{1-\lambda^2}{\lambda^2R_b}\right)
\frac{\overline{M}_f}{M_{\overline f}},
\end{equation}
whereas those corresponding to $\overline{f}\equiv D^{(\ast)-}\pi^+$ 
provide
\begin{equation}
\xi_{\overline{f}}^{(d)}=e^{-i\Theta_{M_{12}}^{(d)}}
\frac{A(\overline{B^0_d}\to \overline{f})}{A(B^0_d\to \overline{f})}=
-\,e^{-i(\phi_d+\gamma)}\left(\frac{\lambda^2R_b}{1-\lambda^2}\right)
\frac{M_{\overline f}}{\overline{M}_f}.
\end{equation}
Here, $R_b$ is the usual CKM factor (see (\ref{Rb-def})), and
\begin{eqnarray}
\overline{M}_f&\equiv&\Bigl\langle f\Bigl|\overline{{\cal O}}_1(\mu)
{\cal C}_1(\mu)+\overline{{\cal O}}_2(\mu){\cal C}_2(\mu)
\Bigr|\overline{B^0_d}\Bigr\rangle\\
M_{\overline{f}}&\equiv&\Bigl\langle\overline{f}\Bigl|{\cal O}_1(\mu)
{\cal C}_1(\mu)+{\cal O}_2(\mu){\cal C}_2(\mu)
\Bigr|\overline{B^0_d}\Bigr\rangle
\end{eqnarray}
are hadronic matrix elements of the following current--current operators:
\begin{equation}
\begin{array}{rclrcl}
\overline{{\cal O}}_1&=&(\overline{d}_\alpha u_\beta)_{\mbox{{\scriptsize 
V--A}}}\left(\overline{c}_\beta b_\alpha\right)_{\mbox{{\scriptsize V--A}}},&
\overline{{\cal O}}_2&=&(\overline{d}_\alpha u_\alpha)_{\mbox{{\scriptsize 
V--A}}}\left(\overline{c}_\beta b_\beta\right)_{\mbox{{\scriptsize V--A}}},\\
{\cal O}_1&=&(\overline{d}_\alpha c_\beta)_{\mbox{{\scriptsize V--A}}}
\left(\overline{u}_\beta b_\alpha\right)_{\mbox{{\scriptsize V--A}}},&
{\cal O}_2&=&(\overline{d}_\alpha c_\alpha)_{\mbox{{\scriptsize V--A}}}
\left(\overline{u}_\beta b_\beta\right)_{\mbox{{\scriptsize 
V--A}}},
\end{array}
\end{equation}
which are completely analogous to the ones we encountered in the
discussion of $\overline{B^0_d}\to D^+K^-$ in 
Subsection~\ref{subsec:ham}. The observables $\xi_f^{(d)}$ and 
$\xi_{\overline{f}}^{(d)}$ allow a {\it theoretically clean} extraction of 
the weak phase $\phi_d+\gamma$ \cite{BDpi}, as the hadronic matrix 
elements $\overline{M}_f$ and $M_{\overline{f}}$ cancel in the following 
combination:
\begin{equation}\label{Prod}
\xi_f^{(d)}\times\xi_{\overline{f}}^{(d)}=e^{-2i(\phi_d+\gamma)}.
\end{equation}
Since $\phi_d$, i.e.\ $2\beta$, can be determined straightforwardly 
with the help of the ``gold-plated'' mode $B_d\to J/\psi\, K_{\rm S}$ (see 
Subsection~\ref{subsec:bpsiks}), we may extract the CKM angle $\gamma$ from 
(\ref{Prod}). As the $\overline{b}\to\overline{u}$ quark-level transition in 
Fig.~\ref{fig:BDpi} is doubly Cabibbo-suppressed by 
$\lambda^2R_b\approx0.02$ with respect to the $b\to c$ transition, 
the interference effects are tiny. However, the branching ratios 
are large, i.e.\ of order $10^{-3}$, and the $D^{(\ast)\pm}\pi^\mp$ states 
can be reconstructed with a good efficiency and modest backgrounds. 
Consequently, $B_d\to D^{(\ast)\pm}\pi^\mp$ decays offer an interesting 
strategy to determine $\gamma$. Experimental feasibility studies can
be found in Refs.~\cite{LHC-Report,Babar-book}.

\section{CP Violation in Charged B Decays}\label{sec:charged}
Since there are no mixing effects present in the charged $B$-meson system, 
non-vanishing $CP$ asymmetries of the kind 
\begin{equation}\label{CP-charged}
{\cal A}_{\mbox{{\scriptsize CP}}}
\equiv\frac{\Gamma(B^+\to f)-\Gamma(B^-\to\overline{f})}{\Gamma(B^+
\to f)+\Gamma(B^-\to\overline{f})}
\end{equation}
would give us unambiguous evidence for ``direct'' $CP$ violation in the 
$B$ system, similarly as $\mbox{Re}(\varepsilon'/\varepsilon)\not=0$ 
does in the neutral kaon system. The $CP$ asymmetries (\ref{CP-charged}), 
which correspond to ${\cal A}_{\rm CP}^{\rm dir}(B_q\to f)$ in 
(\ref{ee6}), arise from interference between decay amplitudes with 
different $CP$-violating weak and $CP$-conserving strong phases. 
Because of the unitarity of the CKM matrix, any $B$-decay amplitude can be 
expressed within the Standard Model in the following way:
\begin{equation}\label{ampl-dec}
A(B^\pm\to f)=|A_1|e^{i\delta_1}e^{\pm i\varphi_1}+
|A_2|e^{i\delta_2}e^{\pm i\varphi_2}.
\end{equation}
Here the $\delta_{1,2}$ are $CP$-conserving strong phases, which are
induced by final-state-interaction (FSI) processes, whereas the 
$\varphi_{1,2}$ are $CP$-violating weak phases, which originate from the 
CKM matrix. Using (\ref{ampl-dec}), we obtain
\begin{equation}\label{ACP-dir-expr}
{\cal A}_{\mbox{{\scriptsize CP}}}=
\frac{-2|A_1||A_2|\sin(\varphi_1-\varphi_2)\sin(\delta_1-\delta_2)}{|A_1|^2+
2|A_1||A_2|\cos(\varphi_1-\varphi_2)\cos(\delta_1-\delta_2)+|A_2|^2}.
\end{equation}
Consequently, a non-vanishing direct $CP$ asymmetry 
${\cal A}_{\mbox{{\scriptsize CP}}}$ requires both a non-trivial strong and
a non-trivial weak phase difference. In addition to the hadronic amplitudes
$|A_{1,2}|$, the strong phases $\delta_{1,2}$ lead to particularly large
hadronic uncertainties in (\ref{ACP-dir-expr}), thereby destroying 
the clean relation to the $CP$-violating weak phase $\varphi_1-\varphi_2$. 

\begin{figure}
\begin{center}
\leavevmode
\epsfysize=3.8truecm 
\epsffile{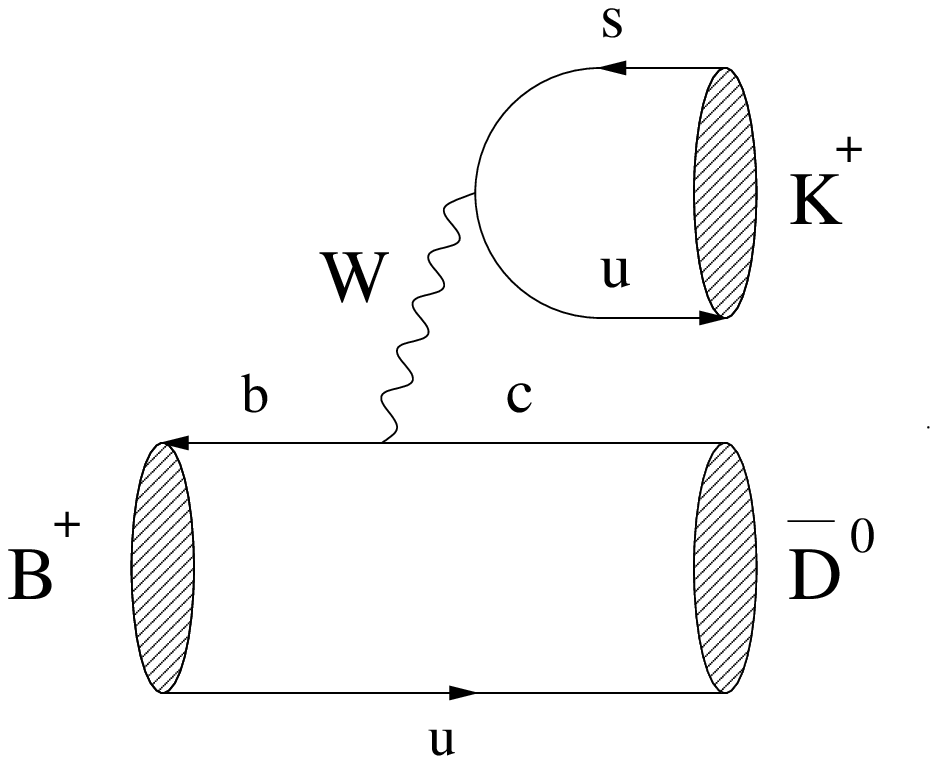} \hspace*{1truecm}
\epsfysize=4.5truecm 
\epsffile{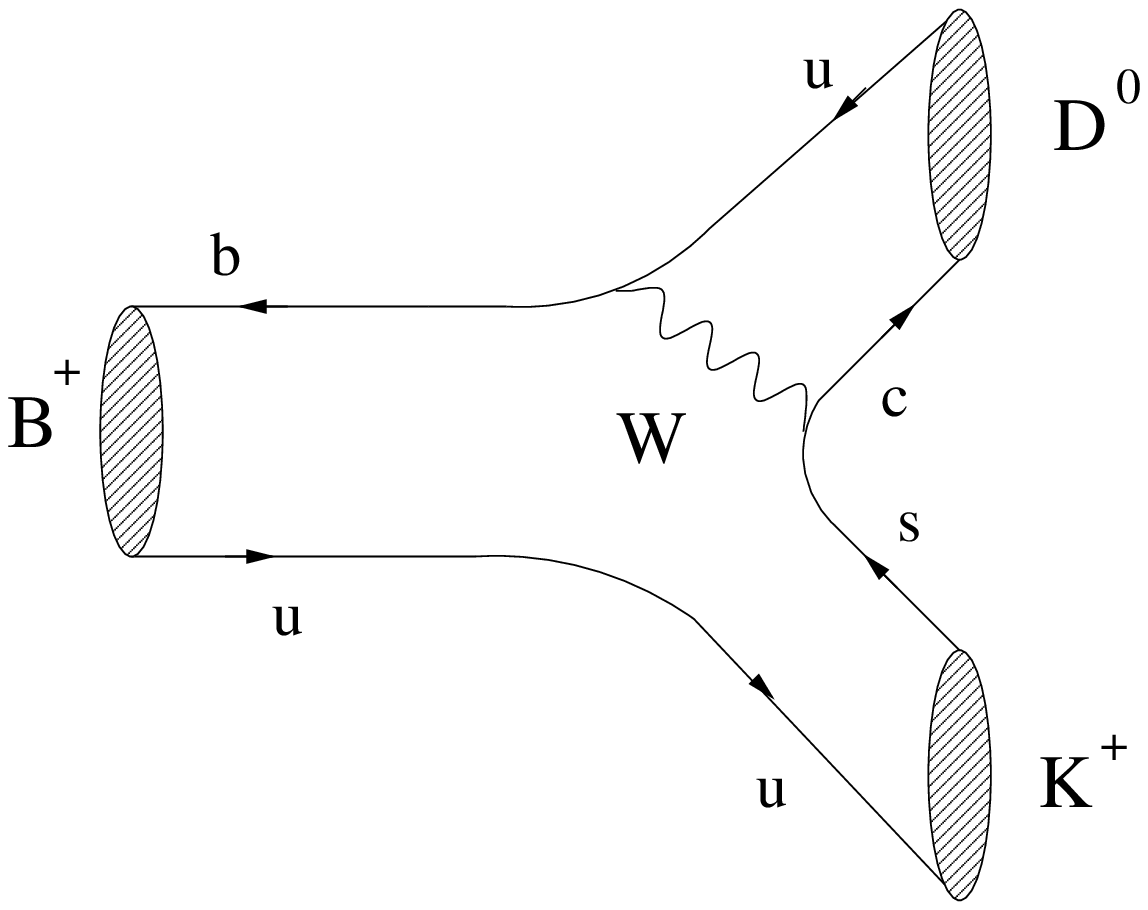}
\end{center}
\vspace*{-0.4truecm}
\caption{Feynman diagrams contributing to $B^+\to \overline{D^0}K^+$ and 
$B^+\to D^0K^+$. }\label{fig:BDK}
\end{figure}

\subsection{Extracting $\gamma$ from $B^\pm\to K^\pm D$ Decays}
\begin{figure}
\centerline{
\epsfysize=2.2truecm
{\epsffile{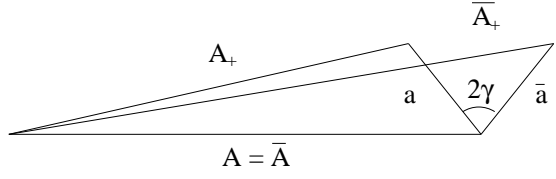}}}
\caption{Triangle relations between charged $B^\pm\to K^\pm D$ decay 
amplitudes.}\label{fig:BDK-triangle}
\end{figure}

An important tool to eliminate the hadronic uncertainties in charged
$B$ decays is given by amplitude relations. The prototype of this 
approach, which was proposed by Gronau and Wyler \cite{gw}, uses 
$B^\pm \to K^\pm D$ decays and allows a {\it theoretically clean} extraction 
of $\gamma$. The decays $B^+\to \overline{D^0}K^+$ and $B^+\to D^0K^+$ are 
pure ``tree'' decays, as can be seen in Fig.~\ref{fig:BDK}. If we make, 
in addition, use of the transition $B^+\to D^0_+K^+$, where $D^0_+$ denotes 
the $CP$ eigenstate of the neutral $D$-meson system with $CP$ eigenvalue 
$+1$,
\begin{equation}\label{ED85}
|D^0_+\rangle=\frac{1}{\sqrt{2}}\left(|D^0\rangle+
|\overline{D^0}\rangle\right),
\end{equation}
we obtain
\begin{equation}
\sqrt{2}A(B^+\to K^+D^0_+)=A(B^+\to K^+D^0)+
A(B^+\to K^+\overline{D^0})
\end{equation}
\vspace*{-0.5truecm}
\begin{equation}
\sqrt{2}A(B^-\to K^-D^0_+)=A(B^-\to K^-\overline{D^0})+
A(B^-\to K^-D^0).
\end{equation}
These relations can be represented as two triangles in 
the complex plane. As we have only to deal with tree-diagram-like topologies,
we have moreover
\begin{eqnarray}
A\,\,\,\equiv\,\,\, A(B^+\to K^+\overline{D^0})&=&A(B^-\to K^-D^0)
\\
a\,\,\,\equiv\,\,\, A(B^+\to K^+D^0)&=&A(B^-\to K^-\overline{D^0})\times
e^{2i\gamma},
\end{eqnarray}
allowing a theoretically clean extraction of $\gamma$ (see 
Fig.~\ref{fig:BDK-triangle}) \cite{gw}. Unfortunately, the 
triangles are very squashed ones, since $a\equiv A(B^+\to K^+D^0)$ is 
colour-suppressed with respect to $A\equiv A(B^+\to K^+\overline{D^0})$:
\begin{equation}
\frac{|a|}{|A|}=\frac{|\overline{a}|}{|\overline{A}|}\approx
\frac{1}{\lambda}\frac{|V_{ub}|}{|V_{cb}|}\times\frac{a_2}{a_1}
\approx0.41\times\frac{a_2}{a_1}\approx0.1.
\end{equation}
Consequently, the $B^\pm\to K^\pm D$ approach is very difficult from
an experimental point of view (see also \cite{ads}). As an alternative,
the decays $B_d\to K^{\ast0}D$ were proposed, where the triangles are
more equilateral \cite{dun}. But all sides are small (colour-suppressed)
so that these decays are also not perfectly suited for the ``triangle''
approach. 

\subsection{Extracting $\gamma$ from $B^\pm_c\to D_s^\pm D$ Decays}
The decays $B_c^\pm\to D_s^\pm D$ are the $B_c$-meson counterparts of 
$B_u^\pm\to K^\pm D$ and also allow an extraction of $\gamma$ 
\cite{masetti}, which relies on the amplitude relations 
\begin{equation}
\sqrt{2}A(B_c^+\to D_s^+D^0_+)=A(B_c^+\to D_s^+D^0)+
A(B_c^+\to D_s^+\overline{D^0})
\end{equation}
\vspace*{-0.5truecm}
\begin{equation}
\sqrt{2}A(B_c^-\to D_s^-D^0_+)=A(B_c^-\to D_s^-\overline{D^0})+
A(B_c^-\to D_s^-D^0),
\end{equation}
with
\begin{eqnarray}
A(B^+_c\to D_s^+\overline{D^0})&=&A(B^-_c\to D_s^-D^0)\\
A(B_c^+\to D_s^+D^0)&=&A(B_c^-\to D_s^-\overline{D^0})\times e^{2i\gamma}.
\end{eqnarray}
At first sight, everything is completely analogous to 
$B^\pm\to K^\pm D$. However, there is an important difference 
\cite{fw}: in the $B_c^\pm\to D_s^\pm D$ system, the amplitude with the 
rather small CKM matrix element $V_{ub}$ is not colour suppressed, while 
the larger element $V_{cb}$ comes with a colour-suppression factor. 
Therefore, we obtain
\begin{equation}
\left|\frac{A(B^+_c\to D_s^+ D^0)}{A(B^+_c\to D_s^+ 
\overline{D^0})}\right|\approx\frac{1}{\lambda}\frac{|V_{ub}|}{|V_{cb}|}
\times\frac{a_1}{a_2}\approx0.41\times\frac{a_1}{a_2}
={\cal O}(1),
\end{equation}
and conclude that the two amplitudes are similar in size. 
Decays of the type $B_c^\pm \to D^\pm D$ 
-- the $U$-spin counterparts of $B_c^\pm\to D_s^\pm D$ -- can be added to 
the analysis, as well as channels, where the $D_s^\pm$- and $D^\pm$-mesons
are replaced by higher resonances. At the LHC, one expects about $10^{10}$ 
untriggered $B_c$-mesons per year of running. Provided there are no serious 
experimental problems, the $B_c^\pm\to D_{(s)}^\pm D$ approach should be 
interesting for the $B$-physics programme of the LHC. From a theoretical 
point of view, it is the ideal realization of the ``triangle'' approach to 
extract $\gamma$. In Section~\ref{sec:BpiK}, we will discuss other strategies
to extract $\gamma$, making use of amplitude relations between $B\to\pi K$ 
decays.

\boldmath
\section{The ``El Dorado'' for Hadron Machines: the $B_s$ 
System}\label{sec:Bs}\unboldmath
Unfortunately, at the $e^+$--\,$e^-$ $B$-factories operating at the 
$\Upsilon(4S)$ resonance, no $B_s$-mesons are accessible. On the other
hand, $B_s$ decays are very promising for hadron machines, where plenty of 
$B_s$-mesons can be produced. There are important differences between the 
$B_d$ and $B_s$ systems:
\begin{itemize}
\item Within the Standard Model, the $B^0_s$--$\overline{B^0_s}$ mixing phase
is negligibly small, 
$\phi_s=-2\lambda^2\eta={\cal O}(0.03)$, whereas $\phi_d=2\beta=
{\cal O}(50^\circ)$.

\item A large mixing parameter $x_s={\cal O}(20)$ is expected in the 
Standard Model (see (\ref{mix-par})), whereas $x_d=0.723\pm0.032$. The 
present experimental lower bound is $\Delta M_s>15.0\,\mbox{ps}^{-1}$,
$x_s>21.3$ (95\% C.L.) \cite{LEPBOSC}.

\item There may be a sizeable width difference 
$\Delta\Gamma_s/\Gamma_s={\cal O}(10\%)$, whereas $\Delta\Gamma_d$ is
negligibly small. The present CDF and LEP average is given by
$\Delta\Gamma_s/\Gamma_s=0.16^{+0.16}_{-0.13}$, $\Delta\Gamma_s/\Gamma_s<
0.31$ (95\% C.L.) \cite{stocchi}.
\end{itemize}

\subsection{$\Delta M_s$ and Constraints in the 
$\overline{\rho}$--$\overline{\eta}$ Plane}\label{Bs-gen}
\begin{figure}
\centerline{\rotate[r]{
\epsfysize=7.5truecm
{\epsffile{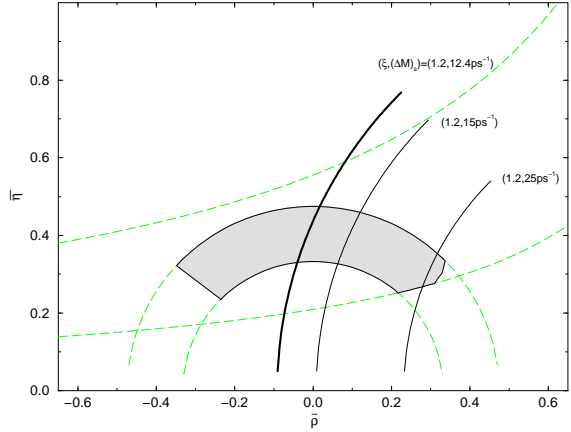}}}}
\caption{The impact of the upper limit $(R_t)_{\rm max}$
on the allowed range in the 
$\overline{\rho}$--$\overline{\eta}$ plane.}\label{fig:UT-constr}
\end{figure}

As we have noted in Subsection~\ref{subsec:CKM-fits}, the mass difference 
$\Delta M_d$ plays an important role to constrain the apex of the unitarity 
triangle shown in \mbox{Fig.\ \ref{fig:UT}\,(a).} 
In particular, it allows us to fix a circle in the 
$\overline{\rho}$--$\overline{\eta}$ plane around $(1,0)$ with radius 
$R_t$. Concerning the theoretical uncertainties, it is -- instead of using 
$\Delta M_d$ separately -- more advantageous to use the ratio
\begin{equation}
\frac{|V_{td}|}{|V_{ts}|}=\xi\sqrt{\frac{m_{B_s}}{m_{B_d}}}
\sqrt{\frac{\Delta M_d}{\Delta M_s}},
\end{equation}
where the $SU(3)$-breaking parameter
\begin{equation}
\xi\equiv\frac{F_{B_s}\sqrt{B_{B_s}}}{F_{B_d}\sqrt{B_{B_d}}}=1.16\pm0.07
\end{equation}
can be determined with the help of lattice or QCD sum rule calculations. 
Interestingly, the presently available experimental lower bound on 
$\Delta M_s$ can be transformed into an upper bound on $R_t$ 
\cite{Buras-Rt}:
\begin{equation}\label{Rtmax}
(R_t)_{\rm max}=1.0\times\xi\times
\sqrt{\frac{10.2/\mbox{ps}}{(\Delta M_s)_{\rm min}}}.
\end{equation}
In Fig.~\ref{fig:UT-constr}, we show the impact of this relation on the 
allowed range in the $\overline{\rho}$--$\overline{\eta}$ plane
\cite{BF-rev}. The strong present lower bound on $\Delta M_s$ excludes 
already a large part in the $\overline{\rho}$--$\overline{\eta}$ plane, 
implying in particular $\gamma<90^\circ$.

\subsection{$\Delta\Gamma_s$ and Untagged Decay Rates}

A non-vanishing width difference $\Delta\Gamma_s$ may allow the extraction 
of $CP$-violating weak phases from the following ``untagged'' $B_s$ rates 
\cite{dunietz}--\cite{DFN}:
\begin{equation}\label{untag-def}
\Gamma[f(t)]\equiv\Gamma(B^0_s(t)\to f)+\Gamma(\overline{B^0_s}(t)
\to f).
\end{equation}
Using (\ref{rates}), we obtain
\begin{eqnarray}
\lefteqn{\Gamma[f(t)]\propto\left[\left(1+\left|\xi_f^{(s)}\right|^2\right)
\left(e^{-\Gamma_{\rm L}^{(s)}t}+
e^{-\Gamma_{\rm H}^{(s)}t}\right)\right.}\nonumber\\
&&\qquad\left.-2\,\mbox{Re}\,\xi_f^{(s)}\left(e^{-\Gamma_{\rm L}^{(s)}t}-
e^{-\Gamma_{\rm H}^{(s)}t}\right)\right],\label{untagged}
\end{eqnarray}
where the $\Delta M_st$ terms cancel. Because of the large mixing parameter 
$x_s$, the $\Delta M_st$ terms arising in ``tagged'' rates oscillate very
rapidly and are hard to resolve. Although it should be not problem to 
accomplish this task at the LHC, studies of untagged rates are interesting 
in terms of 
efficiency, acceptance and purity. Moreover, if $\Delta\Gamma_s$ is sizeable, 
we may extract the observable ${\cal A}_{\Delta\Gamma}(B_s\to f)$ (see 
(\ref{ADGam})) from (\ref{untagged}). Still it is not clear whether
$\Delta\Gamma_s$ is large enough to make this possible. Concerning the
theoretical status of $\Delta\Gamma_s$, next-to-leading order QCD corrections 
were calculated using the heavy-quark expansion \cite{DGamma-cal}; the 
results for $\Delta\Gamma_s$ depend crucially on the size of the relevant 
hadronic matrix elements. The authors of Ref.~\cite{hashimoto} find 
$\Delta\Gamma_s/\Gamma_s=(10.7\pm2.6\pm1.4\pm1.7)\%$, whereas another group 
\cite{rome} gives the smaller value $(4.7\pm1.5\pm1.6)\%$. The difference 
between these results is mainly related to the $B_s$ decay constant 
$F_{B_s}$. Besides ``unquenching'', a better determination of $1/m_b$ 
corrections is very important to reduce the uncertainties of these lattice 
calculations.

\subsection{Pure Tree Decays}
An interesting class of $B_s$ decays is due to $b\to c\,\overline{u}\, s$
quark-level transitions, providing the $B_s$ variant of the 
$B_d\to D^{(\ast)\pm}\pi^\mp$ approach to extract $\gamma+2\beta$
discussed in Subsection~\ref{subsec:BDpi}. Here we have also to deal with
pure ``tree'' decays, where both $B_s^0$- and $\overline{B_s^0}$-mesons
may decay into the same final state $f$. The resulting interference
effects between decay and mixing processes allow a {\it theoretically clean} 
extraction of $\gamma-2\delta\gamma$ from 
\begin{equation}
\xi_f^{(s)}\times\xi_{\overline{f}}^{(s)}=e^{-2i(\gamma-2\delta\gamma)},
\end{equation}
where the $B^0_s$--$\overline{B^0_s}$ mixing phase $-2\delta\gamma$ is 
negligibly small in the Standard Model. It can be probed through 
$CP$-violating effects in $B_s\to J/\psi\,\phi$, as we will see below. 
An interesting difference to the $B_d\to D^{(\ast)\pm}\pi^\mp$ strategy 
is that both decay paths of $B_s^0, \overline{B_s^0}\to f$ are of the same 
order of $\lambda$, thereby leading to larger interference effects.

There are several well-known strategies making use of these features: we may
consider the colour-allowed decays $B_s\to D_s^\pm K^\mp$ \cite{ADK},
or the colour-suppressed modes $B_s\to D^0\phi$ \cite{GL0}. Strategies 
employing ``untagged'' $B_s$ decays were also proposed, where the
width difference $\Delta\Gamma_s$ and the angular distributions of
$B_s\to D_s^{\ast\pm} K^{\ast\mp}$ or $B_s\to D^{\ast0}\phi$ channels play 
a key role \cite{FD2} (see also \cite{LSS}). Recently, strategies using 
``CP-tagged'' $B_s$ decays were proposed \cite{FP}, which require a symmetric 
$e^+$--$e^-$ collider operated at the $\Upsilon(5S)$ resonance. In this 
approach, initially present $CP$ eigenstates $B_s^{\rm CP}$ are employed, 
which can be tagged by making use of the fact that the 
$B_s^0/\overline{B_s^0}$ mixtures have anti-correlated $CP$ eigenvalues 
at $\Upsilon(5S)$. 

\subsection{A Closer Look at $B_s\to J/\psi\,\phi$}\label{BSPSIPHI}
This decay is the $B_s$ counterpart to $B_d\to J/\psi\,K_{\rm S}$ and
offers interesting strategies to extract $\Delta M_s$ and $\Delta\Gamma_s$, 
and to probe $\phi_s=-2\delta\gamma$ \cite{DDF1}. The corresponding
Feynman diagrams are completely analogous to those shown in
Fig.~\ref{fig:BdPsiKS}. Since the final state of $B_s\to J/\psi\,\phi$
is an admixture of different $CP$ eigenstates, we have to use the angular 
distribution of the $J/\psi\to l^+l^-$ and $\phi\to K^+K^-$ decay 
products to disentangle them \cite{DDLR}. The corresponding observables 
are governed by
\begin{equation}\label{Bspsiphi-obs}
\xi^{(s)}_{\psi\phi}\,\propto\, e^{-i\phi_s}
\left[1-2\,i\,\sin\gamma\times{\cal O}(10^{-3})\right],
\end{equation}
where the ${\cal O}(10^{-3})$ factor is an abbreviation for 
$|\lambda_u^{(s)}\tilde A_{\rm pen}^{ut'}|/|\lambda_c^{(s)}
(\tilde A_{\rm cc}^{c'}+\tilde A_{\rm pen}^{ct'})|$ \cite{LHC-Report}.
Since $\phi_s=-2\lambda^2\eta={\cal O}(0.03)$ in the Standard Model, 
there may well be hadronic uncertainties as large as ${\cal O}(10\%)$
in the extraction of $\phi_s$ form the $B_s\to J/\psi[\to l^+l^-]\, 
\phi[\to K^+K^-]$ angular distribution, which may be an important issue
in the LHC era. These hadronic uncertainties can be controlled with the
help of the decay $B_d\to J/\psi\, \rho^0$, which has also some 
other interesting features \cite{RF-ang}. 

\begin{figure}
\centerline{
\epsfysize=4.0truecm
{\epsffile{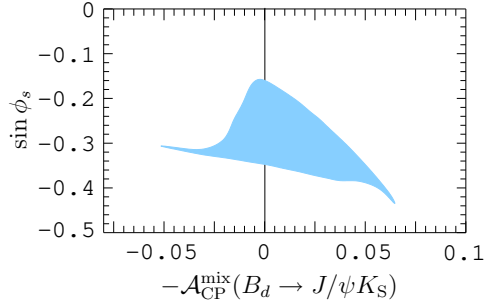}}}
\caption{The allowed region for ${\cal A}_{\rm CP}^{\rm mix}(B_d\to
J/\psi\,K_{\rm S})$ and $\sin\phi_s$ in the SB--LR model.}\label{fig:LR}
\end{figure}

It is an important implication of (\ref{Bspsiphi-obs}) that the 
$CP$-violating effects exhibited by $B_s\to J/\psi\,\phi$ are very small
in the Standard Model, thereby making this channel an interesting probe 
to search for new physics \cite{nir-sil}. A particular scenario for physics
beyond the Standard Model, the symmetrical 
$SU_{\rm L}(2)\times SU_{\rm R}(2)\times U(1)$ model
with spontaneous $CP$ violation (SB--LR) \cite{LR-model1,LR-model2}, 
was considered in Ref.~\cite{BF-LR} to illustrate this feature. Needless to 
note, there are also other scenarios for new physics which are interesting 
in this respect, for example models allowing mixing to a new isosinglet down 
quark, as in $E_6$ \cite{sil}. In the SB--LR model, we obtain the allowed 
region for the mixing-induced $CP$ 
asymmetry in $B_d\to J/\psi\,K_{\rm S}$ and for $\sin\phi_s$ shown in 
Fig.~\ref{fig:LR} \cite{BF-LR}. Here $\phi_s=\phi_s^{\rm SM}+\phi_s^{\rm NP}
=-2\lambda^2\eta+\phi_s^{\rm NP}$, where $\phi_s^{\rm NP}$ originates
from new physics. The quantity $\sin\phi_s$ governs $CP$ violation in 
$B_s\to J/\psi\,\phi$:
\begin{equation}\label{CP1}
\frac{\Gamma(t)-\overline{\Gamma}(t)}{\Gamma(t)+\overline{\Gamma}(t)}=
\left[\frac{1-D}{F_+(t)+D F_-(t)}\right]
\sin(\Delta M_s t)\,\sin\phi_s\,,
\end{equation}
where $\Gamma(t)$ and $\overline{\Gamma}(t)$  denote the time-dependent
rates for decays of initially, i.e.\ at $t=0$, present $B^0_s$- and
$\overline{B^0_s}$-mesons into $J/\psi\,\phi$ final states, respectively,
\begin{equation}\label{dilut}
D\equiv\frac{|A_{\perp}(0)|^2}{|A_0(0)|^2 + |A_{\|}(0)|^2}=
0.1\,\ldots\,0.5,
\end{equation}
is a hadronic factor involving linear polarization 
amplitudes,\footnote{$A_{\perp}$ and $A_0, A_{\|}$ correspond to
$CP$-odd and $CP$-even configurations, respectively.} and 
\begin{equation}\label{Fpm-def}
F_{\pm}(t)\equiv\frac{1}{2}\left[\left(1\pm\cos\phi_s\right)
e^{+\Delta\Gamma_s t/2}+\left(1\mp\cos\phi_s\right)
e^{-\Delta\Gamma_s t/2}\right].
\end{equation}
The range given in (\ref{dilut}) corresponds
to ``factorization'' \cite{DDF1}, and is in agreement
with a recent analysis of the $B_s\to J/\psi\,\phi$ polarization
amplitudes $A_0(0)$, $A_{\|}(0)$, $A_{\perp}(0)$ performed by the
CDF collaboration \cite{CDF-angular}. 

If we look at Fig.~\ref{fig:LR}, we observe that $|\sin\phi_s|$ may be as 
large as 0.4 in the SB--LR model, in contrast to its vanishingly small value 
in the Standard Model. Another interesting feature of this figure is 
the preferred small value for
$-{\cal A}_{\rm CP}^{\rm mix}(B_d\to J/\psi\,K_{\rm S})=\sin\phi_d$. 
As we have noted in Subsection~\ref{subsec:bpsiks}, the first result for
this observable reported by the BaBar collaboration (see (\ref{B-factory})) 
may point towards such a situation. Consequently, it would be very exciting
to measure also (\ref{CP1}), and to explore whether the pattern of small 
$CP$ violation in $B_d\to J/\psi\,K_{\rm S}$ and large $CP$ violation in 
$B_s\to J/\psi\,\phi$ is actually realized in nature \cite{BF-LR}.

\begin{figure}
\centerline{
\epsfysize=5.0truecm
{\epsffile{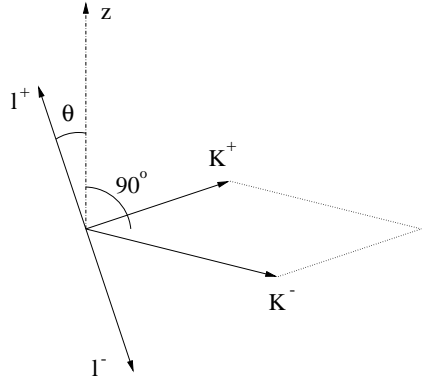}}}
\caption{The kinematics of 
$B_s\to J/\psi\,[\,\to\ l^+l^-]\,\,\phi\,[\,\to\ K^+K^-]$.}\label{fig:kin}
\end{figure}

The $CP$ asymmetry (\ref{CP1}) may provide a ``smoking-gun'' 
signal for new physics, although it does not allow a clean 
determination of $\sin\phi_s$ because of the hadronic parameter
$D$. In order to solve this problem, an angular analysis of the
$B_s\to J/\psi\,[\,\to\ l^+l^-]\,\,\phi\,[\,\to\ K^+K^-]$ decay products 
has to be performed. As the full three-angle distribution is quite
complicated \cite{DDF1}, let us consider here the one-angle distribution: 
\begin{equation}\label{one-angle}
\frac{d \Gamma (t)}{d \cos \Theta} \propto
\left(|A_0(t)|^2 + |A_{\|}(t)|^2\right)\,\frac{3}{8}\,(1 + \cos ^2 \Theta)
+ |A_{\perp}(t)|^2\,\frac{3}{4} \sin^2 \Theta,
\end{equation}
where the kinematics is shown in Fig.~\ref{fig:kin}, corresponding to
the $J/\psi$ rest frame. The one-angle distribution (\ref{one-angle}) 
allows us to extract the observables 
\begin{equation}
P_+(t)\equiv|A_0(t)|^2 + |A_{\|}(t)|^2,\quad P_-(t)\equiv|A_{\perp}(t)|^2, 
\end{equation}
as well as their $CP$ conjugates, thereby providing the $CP$ asymmetries
\begin{equation}\label{ASYM}
\frac{P_\pm(t)-\overline{P}_\pm(t)}{P_\pm(t)+\overline{P}_\pm(t)}=
\pm\,\frac{1}{F_\pm(t)}\,\sin(\Delta M_s t)\,\sin\phi_s.
\end{equation}
On the other hand, untagged data samples are sufficient to determine 
\begin{equation}\label{UNTAG}
P_\pm(t)+\overline{P}_\pm(t)\propto
\left[(1\pm\cos\phi_s)e^{-\Gamma_{\rm L}^{(s)} t}
+(1\mp\cos\phi_s)e^{-\Gamma_{\rm H}^{(s)} t}\right].
\end{equation}
New-physics effects would be indicated by the following features:
\begin{itemize}
\item Sizeable values of the $CP$-violating asymmetries (\ref{ASYM}).
\item The untagged observables (\ref{UNTAG}) depend on two exponentials.
\end{itemize}
In contrast to (\ref{CP1}), these observables do not involve the hadronic
parameter $D$ and allow a clean determination of $\phi_s$. A detailed 
discussion of other strategies to search for new physics with $B_s$ decays 
is given in Ref.\ \cite{DFN}.

\begin{table}[t]
\vspace*{-0.5truecm}
\begin{center}
\caption{$\langle\mbox{BR}\rangle$s in $10^{-6}$ units; the $CP$
asymmetries ${\cal A}_{\rm CP}$ are from CLEO.}\label{tab:BPIK}
\begin{tabular}{ccccc}
\hline
Decay & CLEO & BaBar & Belle & ${\cal A}_{\rm CP}/10^{-2}$\\
\hline
$B_d\to\pi^\mp K^\pm$ & $17.2^{+2.5}_{-2.4}\pm1.2$ & 
$12.5^{+3.0+1.3}_{-2.6-1.7}$ & $17.4^{+5.1}_{-4.6}\pm3.4$ &
$0.04\pm0.16$\\
$B^\pm\to\pi^0K^\pm$ & $11.6^{+3.0+1.4}_{-2.7-1.3}$ & \mbox{} & 
$18.8^{+5.5}_{-4.9}\pm2.3$ & $0.29\pm0.23$\\
$B^\pm\to\pi^\pm K$ & $18.2^{+4.6}_{-4.0}\pm1.6$ & \mbox{} & \mbox{} &
$-0.18\pm0.24$\\
$B_d\to\pi^0 K$ & $14.6^{+5.9+2.4}_{-5.1-3.3}$ & \mbox{} &
$21^{+9.3+2.5}_{-7.8-2.3}$ & \mbox{} \\
\hline
\end{tabular}
\end{center}
\end{table}

\boldmath\section{The Phenomenology of $B\to\pi K$
Decays}\label{sec:BpiK}\unboldmath 
In order to probe $\gamma$, $B\to\pi K$ decays are 
very promising. There are already data on these modes available, 
which triggered a lot of theoretical work. From 1997 until 2000, CLEO 
reported results on $CP$-averaged branching ratios $\langle\mbox{BR}\rangle$; 
in 1999, also studies of $CP$ asymmetries were reported \cite{CLEO}. In 
the summer of 2000, BaBar and Belle announced their first $B\to\pi K$ 
branching ratios \cite{BaBar-bpik,BELLE-bpik}. These results are
collected in Table~\ref{tab:BPIK}.

\subsection{General Remarks}\label{sec:BPIK-gen}

\begin{figure}
%\vspace*{-0.5truecm}
\begin{center}
\leavevmode
\epsfysize=4.0truecm 
\epsffile{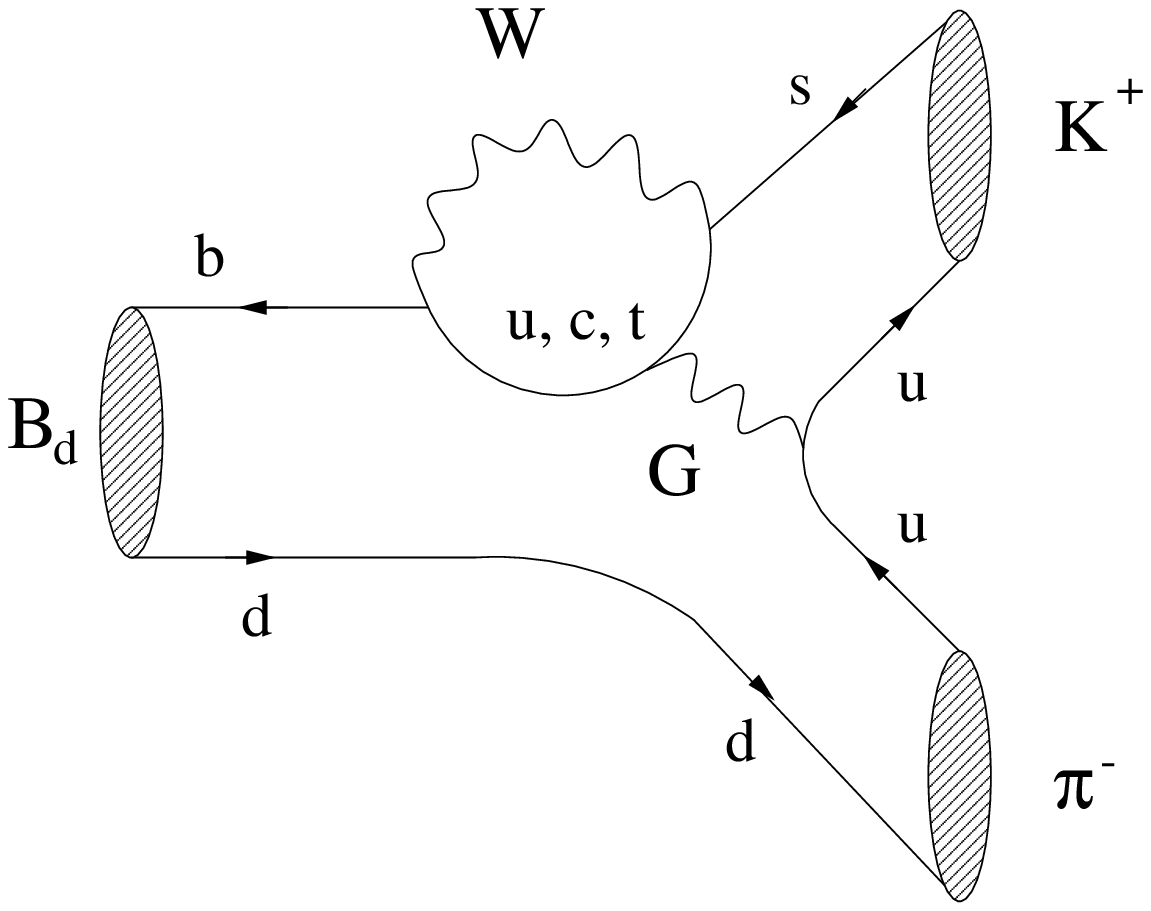} \hspace*{1.4truecm}
\epsfysize=3.5truecm 
\epsffile{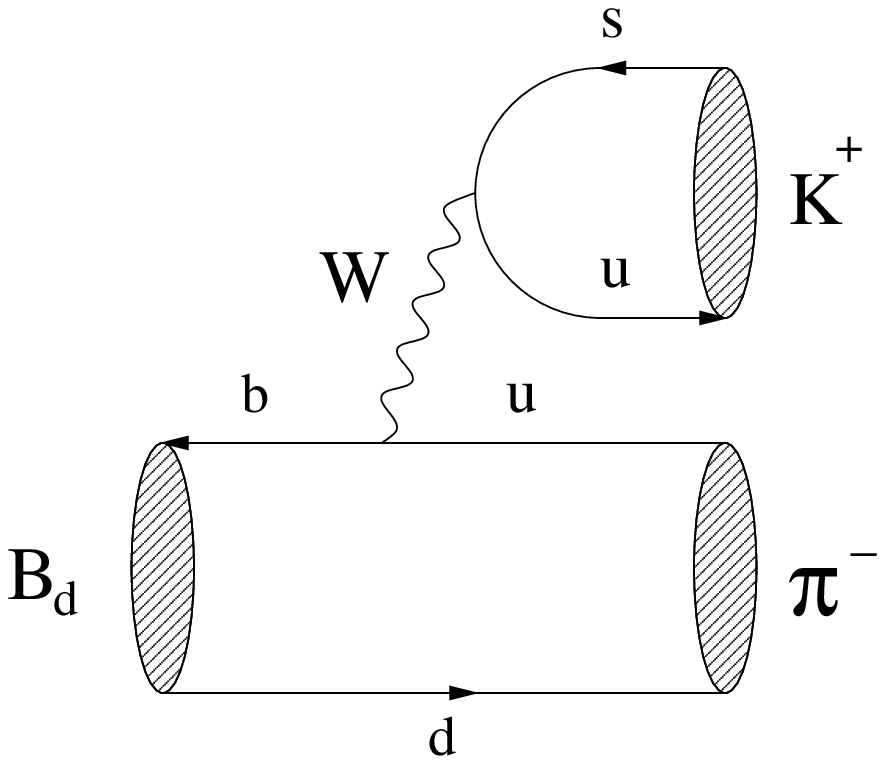}
\end{center}
\vspace*{-0.3truecm}
\caption{Feynman diagrams contributing to 
$B^0_d\to\pi^-K^+$.}\label{fig:BpiK-neutral}
\end{figure}

To get more familiar with $B\to\pi K$ modes, let us consider 
$B^0_d\to\pi^-K^+$. This channel receives contributions from
penguin and colour-allowed tree-diagram-like topologies, as can be 
seen in Fig.~\ref{fig:BpiK-neutral}. Because of the small ratio 
$|V_{us}V_{ub}^\ast/(V_{ts}V_{tb}^\ast)|\approx0.02$, the QCD penguin
topologies dominate this decay, despite their loop suppression. 
This interesting feature applies to all $B\to\pi K$ modes. Because of the 
large top-quark mass, we have also to care about EW penguins. However, in 
the case of $B^0_d\to\pi^-K^+$ and $B^+\to\pi^+K^0$, these topologies 
contribute only in colour-suppressed form and are hence expected to play 
a minor role. On the other hand, EW penguins contribute also in 
colour-allowed form to $B^+\to\pi^0K^+$ and $B^0_d\to\pi^0K^0$, and may 
here even compete with tree-diagram-like topologies. Because of the penguin 
dominance, $B\to\pi K$ modes represent sensitive probes for new-physics 
effects \cite{BPIK-NP,FMat}.

In the Standard Model, the $SU(2)$ isospin symmetry implies 
\begin{displaymath}
\sqrt{2}A(B^+\to\pi^0K^+)+A(B^+\to\pi^+K^0)
\end{displaymath}
\vspace*{-0.5truecm}
\begin{displaymath}
=\sqrt{2}A(B^0_d\to\pi^0K^0)+
A(B^0_d\to\pi^-K^+)
\end{displaymath}
\vspace*{-0.4truecm}
\begin{equation}\label{BPIK-ampl}
=-\left[(T+C)+(P_{\rm ew}+P_{\rm ew}^{\rm C})\right]\propto
\left[e^{i\gamma}+q_{\rm ew}\right],
\end{equation}
where the $(T+C)$ and $(P_{\rm ew}+P_{\rm ew}^{\rm C})$ amplitudes
are due to $($colour-allowed + colour-suppressed$)$ tree-diagram-like and 
EW penguin topologies, respectively. A relation with an analogous phase 
structure holds also for the ``mixed'' $B^+\to\pi^+ K^0$, $B_d^0\to\pi^- K^+$ 
system. So far, strategies to probe $\gamma$ have focused on the following 
systems: $B_d\to\pi^\mp K^\pm$, $B^\pm\to\pi^\pm K$ (``mixed'') 
\cite{BpiK-mixed}--\cite{defan}, and 
$B^\pm\to\pi^0K^\pm$, $B^\pm\to\pi^\pm K$ (``charged'') \cite{NR}. 
Recently, it was pointed out that also the neutral combination 
$B_d\to\pi^\mp K^\pm$, $B_d\to\pi^0 K$ is very promising 
\cite{BF-BpiK1,BF-neut}. 

Interestingly, already $CP$-averaged $B\to\pi K$ branching ratios may lead 
to non-trivial constraints on $\gamma$, which rely on flavour-symmetry 
arguments, involving either $SU(2)$ or $SU(3)$ \cite{GRL,GHLR-SU3}, and 
dynamical assumptions, concerning mainly the smallness of certain rescattering 
effects \cite{FSI}. An example is 
$B^+\to\{\pi^0K^+,\pi^0K^{\ast+},\ldots\}\to\pi^+K^0$. There is still no 
theoretical consensus on the importance of such final-state interaction 
(FSI) processes, although, for instance, the ``QCD factorization'' 
approach~\cite{BBNS} does not suggest 
large effects. However, there are also experimental indicators for large 
FSI effects, for example the decays $B^+\to K^+\overline{K^0}$ 
and $B_d\to K^+K^-$, and methods to include them in the strategies to probe 
$\gamma$ \cite{defan,FSI-strat}. So far, these decays have not been observed 
and the rather strong experimental upper bounds on their $CP$-averaged 
branching ratios are not in favour of dramatic FSI effects 
\cite{BF-neut}, as advocated in some of the papers given in 
Ref.~\cite{FSI}. Let us therefore neglect these effects for a moment.

\subsection{A Simple Example: The ``Mixed'' $B\to\pi K$ System}
Let us illustrate the derivation of the constraints on 
$\gamma$ by considering the ``mixed'' $B^+\to\pi^+ K^0$, $B_d^0\to\pi^- K^+$ 
system \cite{FM,defan}. If we neglect annihilation topologies, penguins 
with internal up quarks and colour-suppressed EW penguins, which are
expected to play a very minor role (but may be enhanced by large FSI
effects; see Subsection \ref{sec:BPIK-gen}), we may write 
\begin{eqnarray}
A(B^+\to\pi^+K^0)&=&\tilde P \,\equiv\,
-\,|\tilde P|e^{i\delta_{\tilde P}}\\
A(B^0_d\to\pi^-K^+)&=&-\left[P+T\right]
\,\equiv\, -\left[-|P|e^{i\delta_P}+|T|e^{i\delta_T}e^{i\gamma}
\right].
\end{eqnarray}
Here $P$ and $T$ describe the penguin and tree-diagram-like
topologies shown in Fig.~\ref{fig:BpiK-neutral}, and $\delta_{\tilde P}$
and $\delta_P$ are $CP$-conserving strong phases. It is convenient to
re-write the $B^0_d\to\pi^-K^+$ decay amplitude as
\begin{equation}
A(B^0_d\to\pi^-K^+)=|P|e^{i\delta_P}\left[1-re^{i\delta}e^{i\gamma}\right],
\end{equation}
where $r\equiv|T|/|P|$ and $\delta\equiv\delta_T-\delta_P$. Consequently,
we obtain the following expressions for the $CP$-averaged decay amplitudes:
\begin{eqnarray}
\left\langle|A(B^\pm\to\pi^\pm K)|^2\right\rangle&=&|\tilde P|^2\\
\left\langle|A(B_d\to\pi^\mp K^\pm)|^2\right\rangle&=&
|P|^2\left[1-2r\cos\delta\cos\gamma+r^2\right].
\end{eqnarray}
Since the $SU(2)$ isospin symmetry implies $|\tilde P|=|P|$,\footnote{Moreover,
$A(B^+\to\pi^+K^0)+A(B^0_d\to\pi^-K^+)=-T$, which corresponds to
(\ref{BPIK-ampl}).} 
we arrive at
\begin{equation}\label{R-param}
R\equiv
\frac{\mbox{BR}(B_d\to\pi^\mp K^\pm)}{\mbox{BR}(B^\pm\to\pi^\pm K)}=
1-2r\cos\delta\cos\gamma+r^2.
\end{equation}
Keeping now the hadronic quantities $\delta$ and $r$ as ``free'' parameters 
yields
\begin{equation}
R\geq\sin^2\gamma,
\end{equation}
which implies $0^\circ\leq\gamma\leq\gamma_0$ $\lor$ 
$180^\circ-\gamma_0\leq\gamma\leq180^\circ$ \cite{FM}, with
\begin{equation}
\gamma_0=\arccos(\sqrt{1-R}).
\end{equation}
This constraint on $\gamma$ is only effective if $R$ is found to be smaller 
than one. In 1997, when CLEO reported the first result on the $CP$-averaged
$B^+\to\pi^+ K^0$, $B_d^0\to\pi^- K^+$ branching ratios, the result
was $R=0.65\pm0.40$. The central value $R=0.65$ would imply 
$\gamma_0=54^\circ$, thereby excluding a large range in the 
$\overline{\rho}$--$\overline{\eta}$ plane. Using the present CLEO data
\cite{CLEO}, we obtain $R=0.95\pm0.28$. Unfortunately, the present 
experimental uncertainties are too large to draw any conclusions and
to decide whether $R<1$. If the parameter $r$ is fixed, for example 
through ``factorization'' \cite{BpiK-mixed}, stronger
constraints on $\gamma$ can be obtained, which are also effective for
$R>1$ \cite{BF-BpiK1,GPY}.

\subsection{The General $B\to\pi K$ Case}\label{subsec:gamma}
In order to constrain $\gamma$, the key quantities are the following ratios:
\begin{eqnarray}
R&\equiv&\frac{\mbox{BR}(B_d\to\pi^\mp K^\pm)}{\mbox{BR}(B^\pm\to\pi^\pm K)}
=0.95\pm0.28\label{CLEO-mix}\\
R_{\rm c}&\equiv&
\frac{2\mbox{BR}(B^\pm\to\pi^0K^\pm)}{\mbox{BR}(B^\pm\to\pi^\pm K)}
=1.27\pm0.47~~\mbox{}\label{CLEO-charged}\\
R_{\rm n}&\equiv&%\frac{1}{2}
\frac{\mbox{BR}(B_d\to\pi^\mp K^\pm)}{2\mbox{BR}(B_d\to\pi^0 K)}
=0.59\pm0.27,\label{CLEO-neut}
\end{eqnarray}
where we have also taken into account the CLEO results reported in 
\cite{CLEO}. If we employ the $SU(2)$ flavour symmetry and
certain dynamical assumptions, concerning mainly the smallness of
FSI effects, we may derive a general parametrization for
(\ref{CLEO-mix})--(\ref{CLEO-neut}) (for an explicit example, see 
(\ref{R-param})) \cite{BF-BpiK1},
\begin{equation}
R_{({\rm c,n})}=R_{({\rm c,n})}(\gamma,q_{({\rm c,n})},r_{({\rm c,n})},
\delta_{({\rm c,n})}),
\end{equation}
where $q_{({\rm c,n})}$ denotes the ratio of EW penguins to ``trees'', 
$r_{({\rm c,n})}$ is the ratio of ``trees'' to QCD penguins, and 
$\delta_{({\rm c,n})}$ is the $CP$-conserving strong phase between 
``tree'' and QCD penguin amplitudes. 

The $q_{({\rm c,n})}$ can be fixed through theoretical arguments: 
in the ``mixed'' $B\to\pi K$ system, we have $q\approx0$, as EW penguins 
contribute only in colour-suppressed form; in the charged \cite{NR} and 
neutral \cite{BF-BpiK1} $B\to\pi K$ systems, $q_{\rm c}$ and $q_{\rm n}$ 
can be fixed through the $SU(3)$ flavour symmetry. On the other hand, the 
$r_{({\rm c,n})}$ can be determined with the help of additional experimental 
information: in the ``mixed'' system, $r$ can be fixed, for example, through 
arguments based on ``factorization'' \cite{BpiK-mixed}, whereas $r_{\rm c}$ 
and $r_{\rm n}$ can be determined from $B^+\to\pi^+\pi^0$ by using again the 
$SU(3)$ flavour symmetry \cite{GRL}. 

At this point, a comment on the FSI effects discussed in 
Subsection~\ref{sec:BPIK-gen} is in order. Whereas the determination of $q$ 
and $r$ as sketched above may be affected by FSI effects, this is {\it not} 
the case for $q_{\rm c,n}$ and $r_{\rm c,n}$, since here $SU(3)$ suffices. 
Nevertheless, we have to assume that $B^+\to\pi^+K^0$ or $B_d^0\to\pi^0K^0$ 
do {\it not} involve a $CP$-violating weak phase, i.e.
\begin{equation}
A(B^+\to\pi^+K^0)=-\,|\tilde P|e^{i\delta_{\tilde P}}
=A(B^-\to\pi^-\overline{K^0}).
\end{equation}
This relation may be affected by rescattering processes as follows:
\begin{equation}
A(B^+\to\pi^+K^0)=-\,|\tilde P|e^{i\delta_{\tilde P}}\left[1+
\rho_{\rm c}\, e^{i\theta_{\rm c}}e^{i\gamma}\right],
\end{equation}
where $\rho_{\rm c}$ is doubly Cabibbo-suppressed and is naively expected to
be negligibly small. In the ``QCD factorization'' approach \cite{BBNS}, 
there is no significant enhancement of $\rho_{\rm c}$ through rescattering 
processes. However, as we have already noted, there is still no theoretical 
consensus on the importance of FSI effects. In the charged $B\to\pi K$ 
strategy to probe $\gamma$, they can be taken into account through $SU(3)$ 
flavour-symmetry arguments and additional data on $B^\pm\to K^\pm K$ 
decays. On the other hand, in the case of the neutral strategy, FSI effects 
can be included in an {\it exact manner} with the help of the mixing-induced 
$CP$ asymmetry ${\cal A}_{\rm CP}^{\rm mix}(B_d\to\pi^0K_{\rm S})$ 
\cite{BF-BpiK1}.

\begin{figure}
\vspace*{-0.4truecm}
\centerline{\rotate[r]{
\epsfysize=7.5truecm
{\epsffile{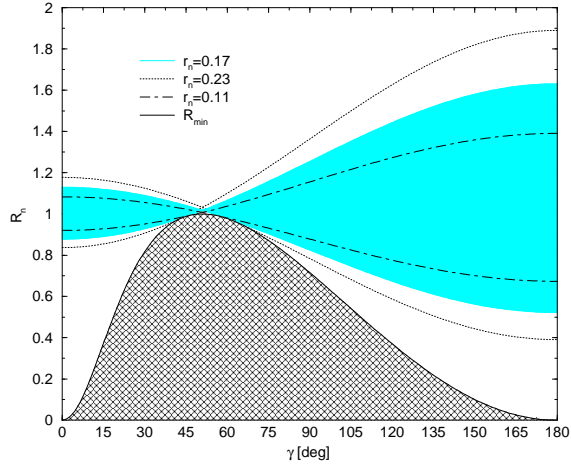}}}}
\vspace*{-0.3truecm}
\caption{The dependence of the extremal values of $R_{\rm n}$ 
%(neutral $B\to\pi K$ system) described by (\ref{const1}) and (\ref{const2}) 
on $\gamma$ for $q_{\rm n}=0.63$.}\label{fig:Rn}
\end{figure}

\begin{figure}
\centerline{\rotate[r]{
\epsfysize=7.9truecm
{\epsffile{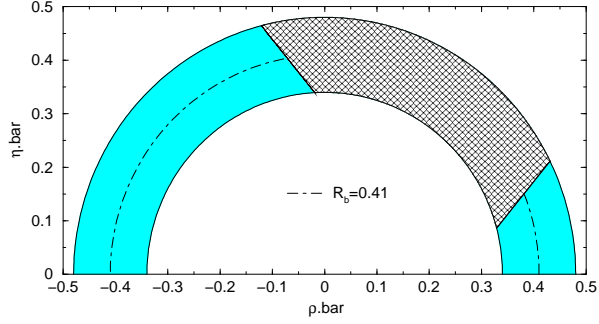}}}}
\vspace*{-0.3truecm}
\caption{The constraints in the $\overline{\varrho}$--$\overline{\eta}$ 
plane for $R_{\rm n}=0.6$
and $q_{\rm n}=0.63 \times[0.41/R_b]$.}\label{fig:rho-eta-bounds}
\end{figure}

In contrast to $q_{({\rm c,n})}$ and $r_{({\rm c,n})}$, the strong phase 
$\delta_{({\rm c,n})}$ suffers from large hadronic uncertainties and is 
essentially unknown. However, we can get rid of $\delta_{({\rm c,n})}$ 
by keeping it as a ``free'' variable, yielding minimal and maximal values 
for $R_{({\rm c,n})}$:
\begin{equation}\label{const1}
\left.R^{\rm ext}_{({\rm c,n})}\right|_{\delta_{({\rm c,n})}}=
\mbox{function}(\gamma,q_{({\rm c,n})},r_{({\rm c,n})}).
\end{equation}
Keeping in addition $r_{({\rm c,n})}$ as a free variable, we obtain 
another -- less restrictive -- minimal value for $R_{({\rm c,n})}$:
\begin{equation}\label{const2}
\left.R^{\rm min}_{({\rm c,n})}\right|_{r_{({\rm c,n})},\delta_{({\rm c,n})}}
=\kappa(\gamma,q_{({\rm c,n})})\sin^2\gamma.
\end{equation}
In Fig.\ \ref{fig:Rn}, we show the dependence of (\ref{const1}) and
(\ref{const2}) on $\gamma$ for the neutral $B\to\pi K$ system.\footnote{The
charged $B\to\pi K$ curves look very similar.}\, Here the crossed region 
below the $R_{\rm min}$ curve, which is described
by (\ref{const2}), is excluded. On the other hand, the shaded region 
is the allowed range (\ref{const1}) for $R_{\rm n}$, arising 
in the case of $r_{\rm n}=0.17$. Fig.~\ref{fig:Rn} allows us to read off 
immediately the allowed region for $\gamma$ for a given value 
of $R_{\rm n}$. Using the central value of the present CLEO result 
(\ref{CLEO-neut}), $R_{\rm n}=0.6$, the $R_{\rm min}$ curve implies 
$0^\circ\leq\gamma\leq21^\circ\,\lor\,100^\circ\leq\gamma
\leq180^\circ$. The corresponding situation in the 
$\overline{\varrho}$--$\overline{\eta}$ plane is shown in 
Fig.~\ref{fig:rho-eta-bounds}, where the crossed region is excluded and 
the circles correspond to $R_b=0.41\pm0.07$. As the theoretical expression 
for $q_{\rm n}$ is proportional to $1/R_b$, the constraints in the 
$\overline{\varrho}$--$\overline{\eta}$ plane are actually more appropriate
than the constraints on $\gamma$.

If we use additional information on $r_{\rm n}$, we may 
put even stronger constraints on $\gamma$. For $r_{\rm n}=0.17$, we obtain
the allowed range $138^\circ\leq\gamma\leq180^\circ$. It is interesting to 
note that the $R_{\rm min}$ curve is only effective for $R_{\rm n}<1$, which 
is favoured by the most recent CLEO data \cite{CLEO}. A similar pattern
is also exhibited by the first Belle results \cite{BELLE-bpik}, 
yielding $R_{\rm n}=0.4\pm0.2$. 

For the central value $R_{\rm c}=1.3$, (\ref{const2}) is not effective and 
$r_{\rm c}$ has to be fixed to 
constrain $\gamma$. Using $r_{\rm c}=0.21$, we obtain 
$87^\circ\leq\gamma\leq180^\circ$. Although it is too early to draw definite 
conclusions, it should be emphasized that the present CLEO results 
on $R_{({\rm c,n})}$ prefer the second quadrant for $\gamma$, 
i.e.\ $\gamma\geq 90^\circ$. Similar 
conclusions were also obtained using other $B\to\pi K$, $\pi\pi$ 
strategies \cite{Hou}. Interestingly, such a situation would be in conflict 
with the standard analysis of the unitarity triangle \cite{AL}, yielding 
$38^\circ\leq\gamma\leq81^\circ$.

The $R_{({\rm c,n})}$ allow us to determine $\cos\delta_{({\rm c,n})}$ 
as functions of $\gamma$, thereby providing also constraints on the strong 
phases $\delta_{({\rm c,n})}$ \cite{BF-neut}.
Interestingly, the present CLEO data are in favour of $\cos\delta_{\rm n}<0$, 
which would be in conflict with ``factorization''. Moreover, they point 
towards a positive value of $\cos\delta_{\rm c}$, which would be in conflict 
with the theoretical expectation of equal signs for $\cos\delta_{\rm c}$ and 
$\cos\delta_{\rm n}$. 

If future data should confirm the ``puzzling'' situation for $\gamma$
and $\cos\delta_{{\rm c,n}}$, which is favoured by the present $B\to\pi K$ 
CLEO data, it may be an indication for new-physics contributions to the EW 
penguin sector, or a manifestation of large flavour-symmetry-breaking 
effects. In order to distinguish between these possibilities, further 
studies are needed. As soon as $CP$ asymmetries in $B_d\to\pi^\mp K^\pm$ 
or $B^\pm\to\pi^0K^\pm$ are observed, 
\begin{equation}
A^{({\rm c,n})}_{\rm CP}=A^{({\rm c,n})}_{\rm CP}
\left(\gamma,q_{({\rm c,n})},
r_{({\rm c,n})},\delta_{({\rm c,n})}\right),
\end{equation}
we may even {\it determine} $\gamma$ and $\delta_{({\rm c,n})}$. 
Here we may also arrive at a situation, where the $B\to\pi K$ observables 
do not provide any solution for these quantities, which would be an 
immediate indication for new physics \cite{FMat}. 

\subsection{Towards the Calculation of $B\to\pi K, \pi\pi$}
Calculations of $B\to\pi K, \pi\pi$ modes are usually based on 
a perturbative picture, where strong phases are obtained from absorptive 
parts of loop diagrams. This approach is referred to as the 
``Bander--Silverman--Soni (BSS) mechanism'' \cite{BSS}. Recently, a 
conceptual improvement of this formalism was presented in Ref.~\cite{BBNS} 
(see also \cite{BL}), where it is argued that there is a heavy-quark 
expansion for non-leptonic $B$-decays into two light mesons, yielding 
transition amplitudes of the structure given in (\ref{QCD-factor}). In 
this approach, soft non-factorizable contributions and FSI effects are 
suppressed by $\Lambda_{\rm QCD}/m_b$. However, the question arises whether 
the $b$-quark mass is large enough to suppress these terms sufficiently. 
Moreover, there are problems due to ``chirally enhanced'' terms, which 
are formally suppressed by $1/m_b$, but are numerically of ${\cal O}(1)$. 
Another ``perturbative'' QCD approach to deal with non-leptonic charmless 
$B$ decays was developed in Ref.\ \cite{PQCD}.

Many calculations of $B\to\pi K$, $\pi\pi$ can be found in the literature 
(see, for instance, \cite{Bpipi-cal}). The results for 
the $CP$-averaged branching ratios are generally in rather good agreement 
with the CLEO data. However, there are two exceptions: $B_d\to\pi^+\pi^-$ 
and $B_d\to\pi^0K$. Concerning the former decay, the calculations favour 
a value at the $7\times 10^{-6}$ level, whereas CLEO finds 
BR$(B_d\to\pi^+\pi^-)=(4.3^{+1.6}_{-1.4}\pm0.5)\times10^{-6}$. Due to 
interference between tree-diagram-like and penguin topologies, the 
theoretical predictions depend on $\gamma$; the CLEO result would be in 
favour of $\gamma>90^\circ$ \cite{Hou}. On the other hand, the first BaBar 
result is BR$(B_d\to\pi^+\pi^-)=(9.3^{+2.6+1.2}_{-2.3-1.4})\times10^{-6}$. 
Concerning $B_d\to\pi^0K$, the calculations favour a value at the 
$5\times10^{-6}$ level, which is essentially independent of 
$\gamma$ and smaller than the CLEO result BR$(B_d\to\pi^0 K)=
(14.6^{+5.9+2.4}_{-5.1-3.3})\times10^{-6}$. Interestingly, the first
Belle result $(21^{+9.3+2.5}_{-7.8-2.3})\times10^{-6}$ is also in favour 
of a large $CP$-averaged $B_d\to\pi^0 K$ branching ratio. Because of the 
large experimental uncertainties, we cannot yet
draw any definite conclusions. However, the experimental situation
should improve significantly in the future. 

\boldmath
\section{The $B_d\to\pi^+\pi^-$, $B_s\to K^+K^-$ 
System}\label{sec:Uspin}\unboldmath
As we have seen in Subsection~\ref{BPIPI-alpha}, $B_d\to\pi^+\pi^-$
is usually considered as a tool to determine $\alpha=180^\circ-\beta-\gamma$. 
Unfortunately, the extraction of $\alpha$ from 
${\cal A}_{\rm CP}^{\rm mix}(B_d\to\pi^+\pi^-)$ is affected by large penguin 
uncertainties, and the strategies to control them through additional data are 
challenging. In this section, we discuss a new approach to employ 
$B_d\to\pi^+\pi^-$: combining this mode with $B_s\to K^+K^-$ -- its
$U$-spin\footnote{$U$-spin is a subgroup of $SU(3)_{\rm F}$, 
relating down and strange quarks to each other.} counterpart -- a 
simultaneous determination of $\phi_d=2\beta$ and $\gamma$ becomes possible 
\cite{RF-BsKK}. In this method, no model-dependent assumptions are required,
and FSI effects, which led to considerable attention in the $B\to\pi K$ 
strategies to probe $\gamma$, do not lead to any problems. The theoretical 
accuracy is hence only limited by $U$-spin-breaking effects. This approach is 
promising for CDF-II ($\left.\Delta\gamma\right|_{\rm exp}=
{\cal O}(10^\circ)$) \cite{wuerth}, and ideally suited for LHCb 
($\left.\Delta\gamma\right|_{\rm exp}={\cal O}(1^\circ)$) \cite{LHC-Report}. 

There are also other interesting strategies to determine $\gamma$ with the 
help of the $U$-spin symmetry: $B_{s(d)}\to J/\psi\, K_{\rm S}$, 
$B_{d (s)}\to D^{\,+}_{d(s)}\, D^{\,-}_{d(s)}$ \cite{RF-BdsPsiK} or 
$B_{s(d)}\to J/\psi\, \eta$ \cite{skands}, as well as $B_{(s)}\to\pi K$ 
\cite{BspiK}. For the ``prehistory'' of $U$-spin arguments in
$B$ decays, the reader is referred to Ref.\ \cite{U-spin}.

\subsection{Extraction of $\beta$ and $\gamma$}
Looking at the diagrams shown in Fig.~\ref{fig:bpipi}, we observe 
that $B_s\to K^+K^-$ is obtained from $B_d\to\pi^+\pi^-$ by interchanging 
all down and strange quarks. The structure of the corresponding decay 
amplitudes is given as follows:
\begin{equation}\label{Bdpipi-ampl0}
A(B_d^0\to\pi^+\pi^-)=\left(1-\frac{\lambda^2}{2}\right){\cal C}
\left[e^{i\gamma}-d\,e^{i\theta}\right]
\end{equation}
\begin{equation}\label{BsKK-ampl0}
A(B_s^0\to K^+K^-)=\lambda\,{\cal C}'\left[e^{i\gamma}+
\left(\frac{1-\lambda^2}{\lambda^2}\right)
d'e^{i\theta'}\right],
\end{equation}
where ${\cal C}'$ and $d'e^{i\theta'}$ take the same form as 
${\cal C}$ and $d\,e^{i\theta}$ (see (\ref{C-DEF}) 
and (\ref{D-DEF})). Using the formalism discussed in 
Subsection~\ref{subsec:CPasym} yields
\begin{eqnarray}
{\cal A}_{\rm CP}^{\rm dir}(B_d\to\pi^+\pi^-)&=&
\mbox{function}(d,\theta,\gamma)\label{ASYM-1}\\
{\cal A}_{\rm CP}^{\rm mix}(B_d\to\pi^+\pi^-)&=&
\mbox{function}(d,\theta,\gamma,\phi_d=2\beta)
\end{eqnarray}
\vspace*{-0.9truecm}
\begin{eqnarray}
{\cal A}_{\rm CP}^{\rm dir}(B_s\to K^+K^-)&=&
\mbox{function}(d',\theta',\gamma)\\
{\cal A}_{\rm CP}^{\rm mix}(B_s\to K^+K^-)&=&
\mbox{function}(d',\theta',\gamma,\phi_s\approx0),\label{ASYM-4}
\end{eqnarray}
where the Standard-Model expectation $\phi_s\approx0$ can be probed
through the decay $B_s\to J/\psi\,\phi$. Consequently, we have four 
observables, depending on six ``unknowns''. However, since 
$B_d\to\pi^+\pi^-$ and $B_s\to K^+K^-$ are related to each other by 
interchanging all down and strange quarks, the $U$-spin flavour symmetry 
of strong interactions implies
\begin{equation}\label{U-spin-rel}
d'e^{i\theta'}=d\,e^{i\theta}.
\end{equation}
Using this relation, the four observables (\ref{ASYM-1})--(\ref{ASYM-4}) 
depend on the four quantities $d$, $\theta$, $\phi_d=2\beta$ and $\gamma$, 
which can hence be determined.

\subsection{Minimal Use of the U-Spin Symmetry}
The $U$-spin arguments can be ``minimized'', if we employ 
$\phi_d=2\beta$ as an input, which can be determined
straightforwardly from $B_d\to J/\psi K_{\rm S}$. The observables
${\cal A}_{\rm CP}^{\rm dir}(B_d\to\pi^+\pi^-)$ and
${\cal A}_{\rm CP}^{\rm mix}(B_d\to\pi^+\pi^-)$ allow us then to 
eliminate the strong phase $\theta$ and to determine $d$ as a function of
$\gamma$. Analogously, ${\cal A}_{\rm CP}^{\rm dir}(B_s\to K^+K^-)$ and 
${\cal A}_{\rm CP}^{\rm mix}(B_s\to K^+K^-)$ allow us to eliminate 
the strong phase $\theta'$ and to determine $d'$ as a function of
$\gamma$. The corresponding contours in the $\gamma$--$d$
and $\gamma$--$d'$ planes can be fixed in a {\it theoretically clean} way.
Using the $U$-spin relation $d'=d$, these contours allow the determination
both of the CKM angle $\gamma$ and of the hadronic quantities 
$d$, $\theta$, $\theta'$. 

\begin{figure}
\vspace*{-1.3truecm}
\centerline{\rotate[r]{
\epsfysize=8.9truecm
{\epsffile{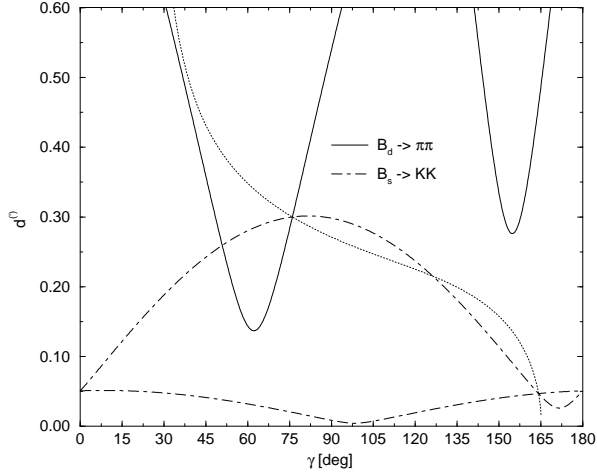}}}}
\caption{The contours in the $\gamma$--$d^{(')}$ planes fixed through the
$CP$-violating $B_d\to\pi^+\pi^-$ and $B_s\to K^+K^-$ observables for an
example discussed in the text.}\label{fig:BsKKcont}
\end{figure}

Let us illustrate this approach by considering the following example: 
\begin{displaymath}
\begin{array}{lcllcl}
{\cal A}_{\rm CP}^{\rm dir}(B_d\to\pi^+\pi^-)&=&+24\%,\,\,& 
{\cal A}_{\rm CP}^{\rm mix}(B_d\to\pi^+\pi^-)&=&+4.4\%,\\ 
{\cal A}_{\rm CP}^{\rm dir}(B_s\to K^+K^-)&=&-17\%,\,\,&
{\cal A}_{\rm CP}^{\rm mix}(B_s\to K^+K^-)&=&-28\%,
\end{array}
\end{displaymath}
which corresponds to the input parameters $d=d'=0.3$, 
$\theta=\theta'=210^\circ$, $\phi_s=0$, $\phi_d=53^\circ$ and 
$\gamma=76^\circ$. In Fig.~\ref{fig:BsKKcont}, the corresponding contours 
in the $\gamma$--$d$ and $\gamma$--$d'$ planes are represented by the solid 
and dot-dashed lines, respectively. Their intersection yields a twofold 
solution for $\gamma$, given by $51^\circ$ and our input value of $76^\circ$. 
The dotted line is related to the quantity
\begin{equation}\label{K-intro}
K\equiv-\left(\frac{1-\lambda^2}{\lambda^2}\right)\left[
\frac{{\cal A}_{\rm CP}^{\rm dir}(B_d\to\pi^+\pi^-)}{{\cal A}_{\rm 
CP}^{\rm dir}(B_s\to K^+K^-)}\right],
\end{equation}
which can be combined with ${\cal A}_{\rm CP}^{\rm mix}(B_s\to K^+K^-)$ 
through the $U$-spin relation (\ref{U-spin-rel}) to fix another contour in 
the $\gamma$--$d$ plane. Combining all contours in Fig.~\ref{fig:BsKKcont} 
with one another, we obtain a single solution for $\gamma$, which is given 
by the ``true'' value of $76^\circ$. As an interesting by-product,
the penguin parameters $d$ and $\theta$, $\theta'$ can be determined
as well, allowing a comparison with theoretical predictions and valuable 
insights into hadronic physics.

\subsection{U-spin-breaking Corrections}
It should be emphasized that the theoretical accuracy of $\gamma$ and
the hadronic parameters $d$, $\theta$, $\theta'$ is only limited
by $U$-spin-breaking effects. In particular, it is not affected by
FSI or penguin effects. Interestingly, the relation (\ref{U-spin-rel}) 
does not receive $U$-spin-breaking corrections within a modernized
version of the ``Bander--Silverman--Soni mechanism'' \cite{BSS}, making 
use -- among other things -- of the factorization hypothesis to estimate 
the relevant hadronic matrix elements \cite{RF-BsKK}. This remarkable
feature strengthens our confidence into the $U$-spin relations used for 
the extraction of $\beta$ and $\gamma$. However, further theoretical studies
of the $U$-spin-breaking effects in the $B_d\to\pi^+\pi^-$, $B_s\to K^+K^-$ 
system, employing, for example, the ``QCD factorization'' approach 
\cite{BBNS}, would be desirable.

Apart from these theoretical considerations, it is also possible to obtain 
experimental insights into $U$-spin breaking. A first consistency check is 
provided by $\theta=\theta'$. Moreover, we may determine the normalization 
factors $|{\cal C}|$ and $|{\cal C}'|$ of the $B^0_d\to\pi^+\pi^-$ and 
$B^0_s\to K^+K^-$ decay amplitudes (see (\ref{Bdpipi-ampl0}) and 
(\ref{BsKK-ampl0})) with the help of the corresponding $CP$-averaged 
branching ratios. Comparing them with the ``factorized'' result
\begin{equation}
\left|\frac{{\cal C}'}{{\cal C}}\right|_{\rm fact}=\,
\left[\frac{f_K}{f_\pi}\right]
\left[\frac{F_{B_sK}(M_K^2;0^+)}{F_{B_d\pi}(M_\pi^2;0^+)}\right]
\left(\frac{M_{B_s}^2-M_K^2}{M_{B_d}^2-M_\pi^2}\right),
\end{equation}
we have another interesting probe for $U$-spin-breaking effects. Moreover,
the $U$-spin relation (\ref{U-spin-rel}) also implies
\begin{equation}
\left[\frac{{\cal A}_{\rm CP}^{\rm dir}(B_s\to K^+K^-)}{{\cal 
A}_{\rm CP}^{\rm dir}(B_d\to\pi^+\pi^-)}\right]=
-\left|\frac{{\cal C}'}{{\cal C}}\right|^2\times
\left[\frac{\mbox{BR}(B_d\to\pi^+\pi^-)}{\mbox{BR}(B_s\to K^+K^-)}\right].
\end{equation}
In order to obtain further insights, the $B_d\to\rho^+\rho^-$, 
$B_s\to K^{\ast+}\,K^{\ast-}$ system would be of particular interest, 
allowing us to determine $\gamma$ together with the mixing phases $\phi_d$ 
and $\phi_s$, and tests of several $U$-spin relations \cite{RF-ang}. 
Further strategies to explore $U$-spin-breaking effects were recently 
discussed in Ref.~\cite{Gronau-U}.

\subsection{Searching for New Physics}
Since penguin processes play an important role in the decays $B_s\to
K^+K^-$ and $B_d\to\pi^+\pi^-$, they -- and the strategy to determine
$\gamma$, where moreover the unitarity of the CKM matrix is employed --
may well be affected by new physics. Interestingly, the Standard Model 
implies a rather restricted region in the space of the $CP$-violating 
observables of the $B_s\to K^+K^-$, $B_d\to\pi^+\pi^-$ system \cite{FMat}. 
A future measurement of observables lying significantly outside of this 
allowed region would be an immediate indication for new physics. 
On the other hand, if the observables should lie within the 
region predicted by the Standard Model, we can extract a value for  
$\gamma$ by following the strategy discussed above. This value for  
$\gamma$ may well be in conflict with other approaches, which would then
also indicate the presence of new physics. 

\section{Remarks on Other Rare Decays}\label{sec:rare}
Let us finally make a few remarks on other ``rare'' $B$ decays, which
occur only at the one-loop level in the Standard Model, and involve 
$\overline{b}\to \overline{s}$ or $\overline{b}\to \overline{d}$ 
flavour-changing neutral-current transitions: $B\to K^\ast\gamma$, 
$B_{s,d}\to \mu^+\mu^-$, $B\to K^\ast\mu^+\mu^-$, 
inclusive decays, etc. Within the Standard Model, these transitions exhibit 
small branching ratios at the $10^{-5}...10^{-10}$ level, do not show 
sizeable $CP$-violating effects, and depend on $|V_{ts}|$ and $|V_{td}|$. 
A measurement of these 
CKM factors from rare decays would be complementary to the one from 
$B^0_{s,d}$--$\overline{B^0_{s,d}}$ mixing. Since they are absent at
the tree level in the Standard Model, rare $B$-decays represent 
interesting probes to search for new physics, and have many other 
interesting aspects. For detailed discussions, the reader is referred 
to the reviews listed in Refs.\ \cite{Buras-lect,Rare}.

\section{Conclusions and Outlook}\label{sec:concl}
The phenomenology of non-leptonic decays of $B$-mesons is very rich and 
has been a field of very active research over the last couple of years. 
These modes provide a very fertile testing ground for the Standard-Model 
description of $CP$ violation, and allow the direct determination of the
angles of the unitarity triangles. Here the goal is to overconstrain
these triangles, which may open a window to the physics beyond the 
Standard Model. As by-products, some strategies allow also the 
determination of interesting hadronic parameters and provide valuable
insights into hadronic physics. Moreover, there are many other exciting 
aspects, for instance the physics potential of certain rare $B$-decays. 
In view of the rich experimental $B$-physics programme of this decade
and the strong interaction between theory and experiment, I have no
doubt that an exciting future is ahead of us. 

\vspace*{0.5truecm}

\begin{flushleft}
{\it Acknowledgements}
\end{flushleft}
I would like to thank the organizers of this NATO Advanced Study Institute
for inviting me to this very interesting meeting in such a pleasant 
environment. Special thanks go to Gustavo Branco, Gui Rebelo and 
Juca Silva--Marcos for their splendid hospitality.

\end{document}